\documentclass[a4paper,11pt]{article}
\usepackage{jheppub} % for details on the use of the package, please see the JINST-author-manual

\usepackage{graphicx}%
\usepackage{multirow}%
\usepackage{amsmath,amssymb,amsfonts}%
\usepackage{amsthm}%
\usepackage{mathrsfs}%
\usepackage[title]{appendix}%
\usepackage{xcolor}%
\usepackage{textcomp}%
\usepackage{manyfoot}%
\usepackage{booktabs}%
\usepackage{algorithm}%
\usepackage{algorithmicx}%
\usepackage{algpseudocode}%
\usepackage{listings}%
\usepackage{xspace}
\usepackage{rotating}
\usepackage{siunitx}
\usepackage{todonotes}
\usepackage{orcidlink}
\usepackage[compat=1.1.0]{tikz-feynman}

\usepackage[parfill]{parskip} % no indent for new paragraphs
\usepackage{lineno}
\DeclareSIUnit{\mrad}{mrad}

\newcommand{\TODO}[1]{\textcolor{red}{TODO: #1}}

\newcommand*{\mrec}{\ensuremath{\text{m}_{\text{recoil}}}\xspace}
\newcommand*{\mh}{\ensuremath{\text{m}_{\text{H}}}\xspace}
\newcommand*{\mZ}{\ensuremath{\text{m}_{\text{Z}}}\xspace}

\newcommand*{\mll}{\ensuremath{\text{m}_{\ell\ell}}\xspace}
\newcommand*{\mjj}{\ensuremath{\text{m}_{\text{jj}}}\xspace}

\newcommand*{\pll}{\ensuremath{\text{p}_{\ell\ell}}\xspace}
\newcommand*{\pjj}{\ensuremath{\text{p}_{jj}}\xspace}

\newcommand{\mff}{m_{\mathrm{f\bar{f}}} \xspace}

\newcommand*{\sigmaZH}{\ensuremath{\sigma_{\text{ZH}}}\xspace} 

\newcommand*{\GeV}{\xspace\ensuremath{\text{GeV}}\xspace}
\newcommand*{\MeV}{\xspace\ensuremath{\text{MeV}}\xspace}
\newcommand*{\ZH}{\ensuremath{\text{ZH}}\xspace}
\newcommand*{\nunuH}{\ensuremath{\nu_e\bar{\nu_e}\text{H}}\xspace}
\newcommand*{\eeH}{\ensuremath{\mathrm{e^{+}e^{-}}\text{H}}\xspace}
\newcommand*{\eeZH}{\ensuremath{\mathrm{e^{+}e^{-}} \rightarrow \text{ZH}}\xspace}
\newcommand*{\eenunuH}{\ensuremath{\mathrm{e^{+}e^{-}} \rightarrow \nu_e\bar{\nu_e}\text{H}}\xspace}
\newcommand*{\eeeeH}{\ensuremath{\mathrm{e^{+}e^{-}} \rightarrow \mathrm{e^{+}e^{-}}\text{H}}\xspace}

\newcommand*{\iso}{\ensuremath{I_{\text{rel}}}\xspace}

\newcommand*{\Z}{\ensuremath{\text{Z}}\xspace} 
\newcommand*{\WW}{\ensuremath{\text{WW}}\xspace} 
\newcommand*{\ZZ}{\ensuremath{\text{ZZ}}\xspace}
\newcommand*{\Zg}{\ensuremath{\text{Z/}\gamma^{*}}\xspace}
\newcommand*{\ttbar}{\ensuremath{\rm t\bar{t}}\xspace}
\newcommand*{\invab}{\ensuremath{ab^{-1}}\xspace}

\newcommand*{\WHIZ}{{\tt WHIZARD3}\xspace}

\newcommand*{\PYTHIAs}{{\tt PYTHIA6}\xspace}
\newcommand*{\PYTHIAe}{{\tt PYTHIA8}\xspace}

\newcommand*{\DELPHES}{{\tt DELPHES}\xspace}
\newcommand*{\KEYfHEP}{{\tt Key4HEP}\xspace}

\newcommand*{\ZeeH}{\ensuremath{\text{Z}(\mathrm{e^{+}e^{-}})\text{H}}\xspace}
\newcommand*{\ZmumuH}{\ensuremath{\text{Z}(\mathrm{\mu^{+}\mu^{-}})\text{H}}\xspace}
\newcommand*{\ZtautauH}{\ensuremath{\text{Z}(\mathrm{\tau^{+}\tau^{-}})\text{H}}\xspace}
\newcommand*{\ZllH}{\ensuremath{\text{Z}(\mathrm{\ell^{+}\ell^{-}})\text{H}}\xspace}
\newcommand*{\ZnunuH}{\ensuremath{\text{Z}(\mathrm{\nu\bar{\nu}})\text{H}}\xspace}
\newcommand*{\ZqqH}{\ensuremath{\text{Z}(\mathrm{q\bar{q}})\text{H}}\xspace}

\newcommand*{\gHZZ}{\ensuremath{g_{\text{HZZ}}}\xspace}
\newcommand*{\gHXX}{\ensuremath{g_{\text{HXX}}}\xspace}
\newcommand*{\sqrts}{\ensuremath{\sqrt{s}}\xspace}
\newcommand*{\sqrtsZH}{\ensuremath{\sqrts = \SI{240}{\GeV}}\xspace}
\newcommand*{\sqrtsTop}{\ensuremath{\sqrts = \SI{365}{\GeV}}\xspace}
\newcommand*{\sqrtsZHTop}{\ensuremath{\sqrts = 240 \text{ and } \SI{365}{\GeV}}\xspace}

\newcommand*{\Hxx}{\ensuremath{\rm H \to X\bar{X}}\xspace}
\newcommand*{\Hbb}{\ensuremath{\rm H \to bb}\xspace}
\newcommand*{\Hcc}{\ensuremath{\rm H \to cc}\xspace}
\newcommand*{\Hgg}{\ensuremath{\rm H \to gg}\xspace}
\newcommand*{\Hss}{\ensuremath{\rm H \to ss}\xspace}
\newcommand*{\Hmumu}{\ensuremath{\rm H \to \mu\mu}\xspace}
\newcommand*{\Htautau}{\ensuremath{\rm H \to \tau\tau}\xspace}
\newcommand*{\HZZ}{\ensuremath{\rm H \to ZZ^{*}}\xspace}
\newcommand*{\HWW}{\ensuremath{\rm H \to WW^{*}}\xspace}
\newcommand*{\HZa}{\ensuremath{\rm H \to Z\gamma}\xspace}
\newcommand*{\Haa}{\ensuremath{\rm H \to \gamma\gamma}\xspace}
\newcommand*{\Hinv}{\ensuremath{\rm H \to \text{inv.}}\xspace}

\arxivnumber{2512.21290} % if you have one

\title{\boldmath Model-independent ZH production cross section at FCC-ee}

\author[a,b]{A. Li,}
\author[c]{J. Eysermans,}
\author[b]{G. Bernardi,}
\author[b]{K. Dewyspelaere,}
\author[d]{M. Selvaggi}
\author[c]{and C. Paus}

\affiliation[a]{Physics Department, Brookhaven National Laboratory,\\
Upton, NY 11973, U.S.A.}
\affiliation[b]{Universit\'e Paris Cit\'e, CNRS, Astroparticule et Cosmologie,\\
F-75013 Paris, France}
\affiliation[c]{Particle Physics Collaboration, Massachusetts Institute of Technology,\\
77 Massachusetts Avenue, Cambridge, MA 02139, U.S.A.}
\affiliation[d]{European Organization for Nuclear Research (CERN),\\
Geneva, Switzerland}
\emailAdd{ang.l@cern.ch}
\emailAdd{jan.eysermans@cern.ch}
\emailAdd{gregorio.bernardi@cern.ch}
\emailAdd{kevin.dewyspelaere@cern.ch}
\emailAdd{michele.selvaggi@cern.ch}
\emailAdd{paus@mit.edu}

\abstract{
This paper presents prospects for measuring the model-independent \ZH production cross section at the FCC-ee using the recoil-mass method at center-of-mass energies of \SI{240}{\GeV} and \SI{365}{\GeV}. Analyses are carried out in the muon, electron, and hadronic decay modes of the associated \Z boson. The event selections rely primarily on the kinematics of the reconstructed \Z decay products, ensuring maximal independence from specific Higgs boson decay modes, while multivariate techniques are employed to further enhance sensitivity.
The statistical interpretation of the leptonic and hadronic final states at 240~GeV, with an integrated luminosity of $10.8~\mathrm{ab}^{-1}$, yields relative precisions of $0.52$\% for the combined leptonic channels and $0.38$\% for the hadronic channel. Their full statistical combination leads to uncertainties of $0.31$\% at 240~GeV and $0.52$\% at 365~GeV, the latter assuming an integrated luminosity of $3.12~\mathrm{ab}^{-1}$. Dedicated statistical tests demonstrate model independence at the level of the obtained precision. This study presents the first consistent and combined analysis of the leptonic and hadronic final states for a model-independent \ZH cross-section measurement at a future lepton collider, using a unified workflow and covering both $\sqrt{s}=240$ and $365~\mathrm{GeV}$. It provides the most precise expected measurement of the \ZH production cross section at future lepton colliders, with the degree of model independence demonstrated within the achieved precision.
}

\begin{document}
\maketitle
\flushbottom

%\clearpage
%\tableofcontents

\section{Introduction} \label{sec:intro}

% TODO
% 1. General reference to FCC: CDR or FSR, or other document? FSR contains latest compatible numbers Ang: (FSR will be good, I already included them in the FCC.bib)
%.2. Should we add the recoil plot here as illustration? E.g. 240/365 for muons, and also for the hadronic channel (e.g. qqH(vvvv)? Depends what is necessary to explain the results and what other plots we have.
% 3. The fact we neglect the fusion events is explained in analysis strategy (together with the vvH, tautautH). It should be obvious to the reader that we measure sigma_ZH, so the fusion events play no role
% 4 We need to reference our 2 reports

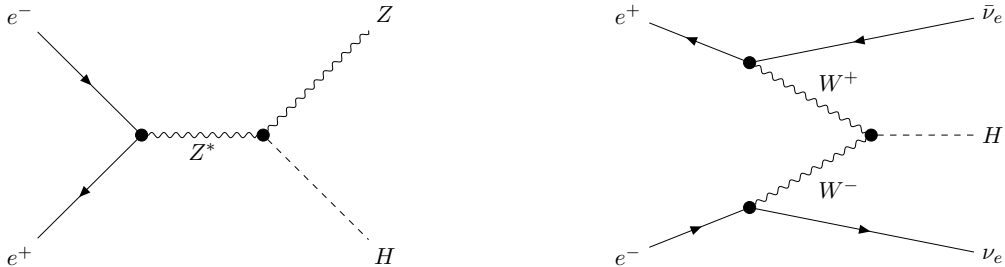
\begin{figure}[!ht]
  \centering
  \scalebox{0.8}{%
    \begin{tikzpicture}
      %========================
      % Left: Higgs-strahlung e+ e- -> ZH
      %========================
      \begin{feynman}[small]
        % incoming leptons
        \vertex (i1) at (-2,  2.0) {\(e^-\)};
        \vertex (i2) at (-2, -2.0) {\(e^+\)};
        % s-channel Z*
        \vertex [dot, inner sep=2pt] (v12) at (0, 0) {};
        % ZH vertex
        \vertex [dot, inner sep=2pt] (vZH) at (2, 0) {};
        % outgoing bosons aligned
        \vertex (oZ) at (4,  2.0) {\(Z\)};
        \vertex (oH) at (4, -2.0) {\(H\)};
    
        \diagram*{
          (i1) -- [fermion] (v12),
          (i2) -- [anti fermion] (v12),
          (v12) -- [boson, edge label'=\(Z^*\)] (vZH),
          (vZH) -- [boson] (oZ),
          (vZH) -- [scalar] (oH),
        };
      \end{feynman}
    
      %========================
      % Right: WW-fusion e+ e- -> ν ν H
      %========================
      \begin{scope}[xshift=10cm] % shift to the right
        \begin{feynman}[small]
          % incoming leptons
          \vertex (ep) at (-2,  2.0) {\(e^+\)};
          \vertex (em) at (-2, -2.0) {\(e^-\)};
    
          % scattering vertices
          \vertex [dot, inner sep=2pt] (vp) at (0,  1.2) {};
          \vertex [dot, inner sep=2pt] (vm) at (0, -1.2) {};
    
          % WW fusion vertex
          \vertex [dot, inner sep=2pt] (vf) at (2, 0) {};
          % outgoing particles aligned
          \vertex (nup) at (4,  2.0) {\(\bar\nu_e\)};
          \vertex (num) at (4, -2.0) {\(\nu_e\)};
          \vertex (h)   at (4,  0.0) {\(H\)};
    
          \diagram*{
            (ep) -- [anti fermion] (vp) -- [anti fermion] (nup),
            (em) -- [fermion] (vm) -- [fermion] (num),
            (vp) -- [boson, edge label=\(W^+\)] (vf),
            (vm) -- [boson, edge label'=\(W^-\)] (vf),
            (vf) -- [scalar] (h),
          };
        \end{feynman}
      \end{scope}
    \end{tikzpicture}
  }
  \caption{Tree-level diagrams of the Higgs production processes at \sqrtsZHTop: the \ZH Higgsstrahlung process (left) and the \nunuH fusion process (right).}
  \label{fig:diagrams}
\end{figure}

The Future Circular Collider (FCC) integrated program foresees an intensity-frontier electron-positron collider (FCC-ee), to be followed by an energy-frontier hadron collider (FCC-hh)~\cite{CDR-FCCee,FCC:2025lpp}. The FCC-ee is designed to operate at several center-of-mass energies ($\sqrt{s}$), from the \Z pole up to the top-quark pair threshold, enabling electroweak measurements with unprecedented precision. In particular, it will operate at \sqrtsZH for three years to study the properties of the Higgs boson, delivering an integrated luminosity of \SI{10.8}{\invab} when summing over the four interaction points. At this energy, the Higgs boson is produced mainly through the Higgsstrahlung process (\eeZH) and, to a lesser extent, via the \WW fusion process (\eenunuH), as illustrated in Fig.~\ref{fig:diagrams}. The \ZZ fusion process (\eeeeH) contributes only marginally. The cross sections for the \eeZH and \WW fusion are 200 and \SI{6.1}{fb} at \sqrtsZH and approximately 125 and \SI{29.2}{fb} at \sqrtsTop, respectively. The expected yields are approximately $2.2\times 10^6$ Higgs bosons produced in the \ZH channel and $6.5\times 10^4$ from \eenunuH fusion. A further five years of running at the top-pair threshold, \sqrtsTop, with an integrated luminosity of \SI{3.12}{\invab}, will yield approximately $3.7\times 10^5$ \ZH events and $9.2\times 10^4$ \nunuH events from \WW fusion. Altogether, nearly three million Higgs bosons will be produced, enabling model-independent determinations of many Higgs couplings with per-mille-level precision and of the Higgs boson mass to a few\MeV~\cite{eysermans_2025_jfb44-s0d81,DelVecchio:2025gzw,Selvaggi:2025kmd}.

One of the flagship measurements of the FCC-ee Higgs physics program is the determination of the total \ZH production cross section, $\sigmaZH$, in a model-independent manner, providing direct access to the absolute coupling $\rm\gHZZ$. This measurement is made possible at an electron-positron collider thanks to the complete knowledge of the initial-state center-of-mass energy. In contrast, at hadron colliders the initial state is ill-defined due to parton distribution functions and non-perturbative effects, since only a fraction of each proton participates in the hard interaction, and this fraction cannot be measured on an event-by-event basis. At lepton colliders, one can exploit the recoil-mass technique, in which events are selected by identifying the associated \Z boson decaying into a pair of fermions (either leptons or quarks, the latter subsequently hadronizing). The mass computed from the four-momentum of the system opposite to the reconstructed \Z boson, $\mrec$, corresponds to the Higgs boson mass, up to detector resolution and radiative effects:
\begin{equation}\label{Equ::recoil::Recoil}
    \rm \mrec^2 = (\sqrt{s} - E_{f\bar{f}})^2 - p^2_{f\bar{f}} = s - 2E_{f\bar{f}}\sqrt{s} + m^2_{f\bar{f}}. 
\end{equation}

Here, $\rm E_{f\bar{f}}$, $\rm p_{f\bar{f}}$ and $\rm m_{f\bar{f}}$ denote the energy, momentum, and mass of the reconstructed difermion system, respectively. $\sqrt{s}$ denotes the nominal average center-of-mass energy, precisely determined using accelerator instrumentation and independently measured in data using radiative-return events to the $\rm Z$ boson. It is set to 240 and 365~GeV for the two FCC-ee data-taking energies considered. Once the leptons or hadrons originating from the associated 
\Z boson are identified, the event selection becomes independent of the Higgs boson decay products, making the measurement model independent. While absolute model independence cannot be guaranteed in a strict sense, sensitivity to the Higgs boson decay mode is minimized, and potential biases are explicitly addressed. Dedicated bias tests are used to ensure that any remaining model dependence is controlled and remains within the quoted measurement uncertainties.

Summed over both center-of-mass energy points, nearly 2.57 million \ZH events are expected at FCC-ee, corresponding to an ultimate statistical limit of about $0.06$\% on a common \ZH signal strength and of approximately $0.03$\% on $\rm\gHZZ$.
This illustrates the unprecedented sensitivity offered by the FCC-ee program. Knowledge of $\rm\gHZZ$ provides a model-independent normalization that allows all other Higgs couplings to be determined through measurements of $\sigmaZH \times \mathcal{B}(\Hxx)$, which scale as:

\begin{equation}
    \sigmaZH \times \mathcal{B}(\Hxx) \propto \frac{\gHZZ^{2}\, \gHXX^{2}}{\Gamma_H}.
\end{equation}

In combination with a direct determination of the Higgs width ($\rm \Gamma_H$) from \HZZ decays, this measurement enables the extraction of all other Higgs couplings in a model-independent manner. In practice, the Higgs couplings are derived through a comprehensive global fit that optimally combines all available measurements at FCC-ee. The inclusion of results obtained at \sqrtsTop in the global fit significantly improves, and ultimately drives, the precision on $\rm \Gamma_H$, owing to the interplay between VBF production with \Hbb decays and \ZH production with \HWW decays. In this approach, a precision of 0.78\% on $\rm \Gamma_H$ is achievable at FCC-ee~\cite{FCC:2025lpp}.

In this paper, the recoil-mass method is applied to events in which the associated \Z boson decays to electrons, muons, or hadrons. Leptonic final states provide a clean reconstruction thanks to the excellent momentum resolution for leptons, but suffer from small branching ratios. By contrast, the hadronic final state benefits from a branching ratio roughly twenty times larger than each of the electron and muon channels, at the cost of poorer kinematic resolution, larger backgrounds, and increased ambiguity due to overlaps with Higgs boson decay products. Decays of the associated \Z boson to $\tau$ leptons or neutrinos are not used as explicit recoil channels, as the direct reconstruction of the associated \Z boson is degraded by the presence of undetected neutrinos. They are nevertheless included in the signal definition, in particular for the evaluation of Higgs-decay-dependent selection efficiencies and possible biases in the model-independence tests.

This study represents the first consistent implementation of an analysis combining \Z boson decays to electron, muon, and hadronic final states, using a common selection for the electron and muon channels and an orthogonal, well-defined selection for the hadronic channel, combined through a rigorous statistical procedure. The combination improves the statistical precision and provides a consistency test across channels with different experimental sensitivities. Dedicated statistical tests validate the degree of model independence at the level of the quoted combined precision.

Similar studies have been carried out for other proposed electron-positron colliders. At the Linear Collider Facility (LCF), an analysis was performed at a center-of-mass energy of \SI{250}{\GeV} for a total integrated luminosity of \SI{2.7}{\invab} under polarized beam conditions that enhance or suppress specific signal and background processes. Separate investigations of the leptonic~\cite{PhysRevD.94.113002} and hadronic~\cite{Thomson:2015jda} channels were subsequently rescaled to the latest LCF running scenario, yielding a total uncertainty of 0.62\%~\cite{LinearColliderVision:2025hlt}. In these studies, the visible and invisible Higgs decay modes are treated in separate, independently optimized analyses, in contrast to the integrated strategy pursued in this work. At the proposed Circular Electron-Positron Collider (CEPC), dedicated studies have been performed for the leptonic channels, while the hadronic contribution was extrapolated from previous results for linear colliders, leading to a projected total uncertainty of 0.5\% for \SI{5.6}{\invab} of integrated luminosity at \sqrtsZH~\cite{An_2019}.

This paper is organized as follows. Section~\ref{sec:AnalysisStrategy} provides an overview of the analysis strategy, followed by a description of the event generation and detector response in Section~\ref{sec:Samples}. The event selection is detailed in Section~\ref{sec:EventSelection}, while Section~\ref{sec:SignalExtraction} presents the final optimization and signal extraction. The results are given in Section~\ref{sec:results}, followed by the assessment of model independence in Section~\ref{sec:Model-independent}, and Section~\ref{sec:Conclusion} concludes the paper.

\section{Analysis Strategy}\label{sec:AnalysisStrategy}

This paper targets the \ZeeH, \ZmumuH, and \ZqqH final states. The analyses follow a common strategy based on the recoil-mass method. The leptons or jets from the associated \Z decay are first identified, and events are then selected to suppress backgrounds (\WW, \ZZ, and \Zg processes) using only their kinematic properties. In the leptonic final states, tight kinematic selections are possible thanks to the excellent momentum resolution, while in the hadronic final state a looser selection is adopted to minimize model dependence.  

After the baseline selection, multivariate techniques (MVA) are employed to further enhance sensitivity. In the leptonic final states, the signal yield and its uncertainty are extracted from a maximum-likelihood fit to the recoil mass distribution, performed in two regions of the MVA discriminator. In the hadronic final state, the larger event yields and looser selection enable a two-dimensional maximum-likelihood fit to the recoil and jet-pair mass distributions, also separated into two regions of the MVA output. Orthogonality among the final states is ensured by selecting or vetoing leptons in the event selection, which makes a consistent combination of the three final states possible.

In addition to the three targeted final states, \ZtautauH and \ZnunuH events can also pass the selection criteria and are therefore included in the signal definition. Higgs production through \nunuH and \eeH fusion is also included, but the selection efficiency for these processes is very low and the total number of events after the selection is negligible. Consequently, the signal considered in this analysis corresponds to the total \ZH production rate.

\section{Event Generation and Detector Response}\label{sec:Samples}

This study relies on samples produced with multiple Monte Carlo generators, with the detector response modeled using fast simulation.

\subsection{Event generation}

The Monte Carlo samples used in this analysis were generated at \sqrtsZHTop. They are normalized to integrated luminosities of 10.8 and \SI{3.12}{\invab}, respectively, corresponding to the four-interaction-point layout of FCC-ee.

A beam energy spread of 0.185\% (0.221\%), equivalent to \SI{222}{\MeV} (\SI{403}{\MeV}), is applied to both beams. The interaction vertex is smeared according to realistic conditions described in the FCC Feasibility Study Report~\cite{FCC:2025lpp}.

Event generation is performed with \WHIZ~\cite{Kilian:2007gr} interfaced to \PYTHIAs~\cite{Sjostrand:2006za} for the \ZH signal and for the $\mathrm{Z}/\gamma^*$ fermion-pair, two-photon, $\mathrm{e}\gamma$, and vector-boson-fusion backgrounds, while the diboson (\WW and \ZZ) samples and the $\mathrm{t\bar{t}}$ samples at $\sqrt{s} = 365$~GeV are generated with \PYTHIAe~\cite{Sjostrand:2014zea} standalone.
In the \WHIZ samples, the hard-scattering process is simulated at tree level, with initial-state radiation and beam-energy spread handled in \WHIZ. \PYTHIAs is used for parton showering, hadronization, and the decays of the Higgs boson, with its internal initial-state radiation disabled to avoid double counting. In the diboson and \ttbar samples, \PYTHIAe provides the matrix elements as well as the subsequent parton showering and hadronization. No fixed-order NLO QCD or electroweak corrections are included for the hard production processes.

The signal process \eeZH is generated using dedicated samples for each Z boson decay mode. The \Z boson is allowed to decay to electron, muon, tau, quark, or neutrino pairs, while the Higgs boson is treated inclusively. The $\mathrm{e^+e^-} \to \mathrm{e^+e^-H}$ and $\mathrm{e^+e^-} \to \nu\bar{\nu}\mathrm{H}$ samples are generated as full $2 \to 3$ processes and therefore include the ZZ- and WW-fusion contributions, respectively, together with their interference with ZH. The $\mathrm{Z} \to \tau^+\tau^-$ and $\mathrm{Z} \to \nu\bar{\nu}$ samples are not used as explicit recoil channels, but are included in the signal modeling and in the model-independence tests.

The background samples include diboson production ($\rm W^+W^-$ and $\rm ZZ$), fermion-pair production ($\rm e^+e^- \to Z/\gamma^* \to f\bar{f}$), two-photon interactions, electron–photon scattering ($\rm e^\pm\gamma \to e^\pm Z$), and vector-boson-fusion production of $\rm e^+e^- \to \nu_e\bar{\nu}_e Z$. The $\rm W^+W^-$ sample is generated with inclusive $\rm W$-boson decays. The neutral-current diboson sample, denoted as $\rm ZZ$, is generated as $\rm e^+e^- \to (\gamma^*/Z)(\gamma^*/Z)$, with inclusive decays of the two neutral bosons. It therefore includes the $\rm ZZ$, $\rm Z\gamma^*$, and $\rm \gamma^*\gamma^*$ contributions and their interference within the double-resonant approximation implemented in \PYTHIAe{}. The fermion-pair samples are restricted to two-fermion hard-process final states, with initial- and final-state photon radiation included as described above. The $\rm e^\pm\gamma \to e^\pm Z$ samples describe the quasi-real photon-induced component of $\rm e^+e^- \to e^+e^-Z$ through the equivalent photon approximation. At \sqrtsTop, \ttbar production is kinematically accessible, but dedicated samples, generated with \PYTHIAe, show that its contribution is negligible in the selected event phase space.

All events are normalized to their respective production cross sections as provided by the event generators. Uncertainties in the background cross sections are included as nuisance parameters in the statistical model, as discussed in Section~\ref{subsec:systematics}. The total dataset spans event counts from a few million for signal and rare processes to several hundred million for diboson and fermion-pair processes, with cross sections ranging from $\mathcal{O}(10^{-2})$ to $\mathcal{O}(10)~\mathrm{pb}$.

\subsection{Detector response}

Detector simulation and response are performed with \DELPHES~\cite{deFavereau:2013fsa}, a fast simulation package that applies parameterized resolutions and efficiencies to generator-level particles to emulate a realistic detector performance. The output consists of particle-flow candidates that are directly used in the analysis. The full chain of event generation, simulation, and reconstruction is integrated within the \KEYfHEP software framework~\cite{key4hep}. A right-handed coordinate system is used, with its origin at the nominal interaction point at the detector center: the $z$-axis is aligned with the beam pipe, the $x$-axis points towards the center of the ring, and the $y$-axis points upwards. Spherical coordinates $(r,\theta,\phi)$ are defined with $\theta$ as the polar angle from the $+z$ direction and $\phi$ the azimuthal angle around the $z$-axis.

Among the detector concepts considered for FCC-ee, this analysis employs a modified version of the Innovative Detector for Electron-positron Accelerator (IDEA)~\cite{IDEA1,IDEA2}, which serves as the default detector model for FCC-ee simulation studies. The IDEA design features a five-layer silicon pixel vertex detector, surrounded by a lightweight drift chamber with up to 112 sensitive layers, providing quasi-continuous tracking with excellent performance. Both tracking systems are contained within a thin \SI{2}{\tesla} solenoidal magnetic field.

Outside the solenoid, a dual-readout calorimeter provides measurements of electromagnetic and hadronic particles, while the detector is enclosed by $\mu$-RWELL muon chambers, a technology combining Resistive Plate Chambers and Gas Electron Multipliers. The overall detector layout, characterized by a large tracking volume and a low material budget in the tracking region, is optimized for excellent momentum resolution.

For this study, a slightly modified version of IDEA is adopted, in which a crystal electromagnetic calorimeter~\cite{Lucchini:2020bac} is placed in front of the hadronic dual-readout calorimeter, resulting in a significantly improved electron energy resolution. In the simulation, the track parameter resolutions of electrons are scaled by a factor of $1.25$ with respect to those of muons, i.e.\ the track covariance matrix is scaled by $1.25^2$, to reproduce the degradation of the electron momentum resolution from bremsstrahlung studied in full simulation~\cite{eperf}. At the typical lepton momenta of this analysis, $\rm p \simeq 50$ and $\SI{85}{\GeV}$ at \sqrtsZHTop, respectively, the momentum resolution is $\rm \sigma_p/p \simeq 0.15\%$ ($0.19\%$) for muons (electrons) at \sqrtsZH and $0.25\%$ ($0.31\%$) at \sqrtsTop. The parameterized calorimeter response yields a relative jet energy resolution of approximately 4\% for jets with an energy of \SI{50}{\GeV}, representative of the typical jet energies at \sqrtsZH. Charged particles are reconstructed with full efficiency within $\lvert \cos\theta \rvert < 0.988$, and electrons and muons with $\rm E > \SI{2}{\GeV}$ are identified with an efficiency of 99\%. Lepton identification is based on the generator-level particle identity in the fast simulation, so the misidentification of hadrons as leptons is not modeled, and no fake-lepton contribution is included. Non-prompt leptons from semileptonic heavy-flavor hadron decays inside jets are simulated and are suppressed by the isolation requirement of Section~\ref{para:objects} and the kinematic selection of Section~\ref{para:selection}.

\section{Event Selection}\label{sec:EventSelection}

The analyses utilize the recoil-mass method, starting by identifying the leptons or jets from the associated \Z decay (see Section~\ref{para:objects}). A subsequent kinematic selection is applied to suppress backgrounds (see Section~\ref{para:selection}), dominated by \WW, \ZZ, and \Zg, with rare processes included as well. Thanks to the excellent momentum resolution, tighter selections can be applied in the leptonic final states, while in the hadronic final state a looser selection is adopted to minimize dependence on the Higgs decay mode.

Unless stated otherwise, the same selection strategy is used at both \sqrtsZHTop, with electrons and muons treated identically and referred to collectively as leptons.  

\subsection{Object selection} \label{para:objects}

Before applying kinematic selections to define the signal region, objects with the highest probability of originating from the associated \Z boson are identified. In both the leptonic and hadronic final states, the \Z candidate is chosen as the fermion pair that minimizes
\begin{equation}\label{eq:chi2}
    \rm \chi^2 = A \times (\mff - \mZ)^2 + B \times (\mrec - \mh)^2,
\end{equation}
where $\rm \mff$ is the invariant mass of the lepton pair ($\rm \mll$) or jet pair ($\rm \mjj$), $\mrec$ the recoil mass against it, $\mZ=\SI{91.2}{\GeV}$, and $\mh=\SI{125}{\GeV}$. Only the ratio $\rm A/B$ affects the minimization, and the coefficients are set separately for the two final states as described below.

For the leptonic final states, events are required to contain at least two opposite-sign (OS) and same-flavor leptons with momentum $\rm p > \SI{20}{\GeV}$, in order to suppress low-energy leptons from soft radiation or leptonic $\tau$ decays. To further reduce backgrounds, mainly from non-prompt leptons produced in semileptonic heavy-flavor hadron decays, at least one lepton must be isolated with $\rm \iso < 0.25$, where the relative cone isolation $\rm \iso$ is defined as the scalar sum of all particle-flow candidates' momenta within a cone of radius $\rm \Delta R < 0.5$ around the direction of the lepton, divided by the lepton momentum. The variable $\rm \Delta R$ is defined as $\rm \Delta R = \sqrt{(\Delta \theta)^2 + (\Delta \phi)^2}$, where $\theta$ and $\phi$ are the polar and azimuthal angle, respectively. If more than one lepton pair can be formed, the pair minimizing Eq.~\eqref{eq:chi2} with $\rm A=0.6$ and $\rm B=0.4$ is selected. These values reflect the different resolutions of the lepton-pair invariant mass and the recoil mass, and have been further optimized to minimize the selection dependency across different Higgs boson decay modes. To suppress \Hmumu decays, lepton pairs with mass within \SI{3}{\GeV} of $\mh$ are excluded from the $\chi^2$ minimization.

In the hadronic final state, events already selected in the leptonic final states are vetoed. As this veto is defined by the leptonic selection, which requires an isolated lepton, non-prompt leptons from semileptonic heavy-flavor decays inside jets that fail the isolation requirement do not remove events from the hadronic final state. Since hadronic events contain a wide variety of objects depending on the Higgs decay, several jet clustering algorithms are evaluated per event to improve the robustness of the \Z candidate reconstruction against different Higgs decay topologies, additional QCD radiation, and possible jet merging or acceptance effects. The algorithms considered are exclusive clustering with $\rm N=2$, 4, and 6 jets, and inclusive clustering. The exclusive Durham algorithm~\cite{Catani:1991hj} is used, a sequential spherical $\rm k_t$ algorithm that iteratively combines the two closest particles until the desired jet multiplicity is reached. Inclusive clustering is performed with the $\rm k_t$ algorithm using a radius parameter of 0.6 in the rapidity--azimuth plane. Both algorithms are implemented in the \texttt{FastJet} package~\cite{Fastjet}. Prior to clustering, identified photons and leptons are removed. Jets with $\rm p < \SI{5}{\GeV}$ are discarded, and jet collections are required to contain at least two jets. For each remaining clustering scheme, Eq.~\eqref{eq:chi2} is evaluated for all jet-pair combinations with equal weights, $\rm A=B=1$, and the jet pair with the lowest $\chi^2$ across all schemes is selected as the \Z candidate, allowing different clustering schemes to be exploited depending on the event topology. Unlike the leptonic case, $\mjj$ and $\mrec$ are here both dominated by the jet energy resolution, so no relative weighting is applied to both terms. The ratio $\rm A/B$ was nevertheless scanned as for the leptonic final states, and the selection efficiency and its variation across Higgs boson decay modes were found to be insensitive to it.

\begin{figure}[!ht]
\centering
\hspace*{-7pt} % align both row of images vertically
\includegraphics[width=0.325\linewidth]{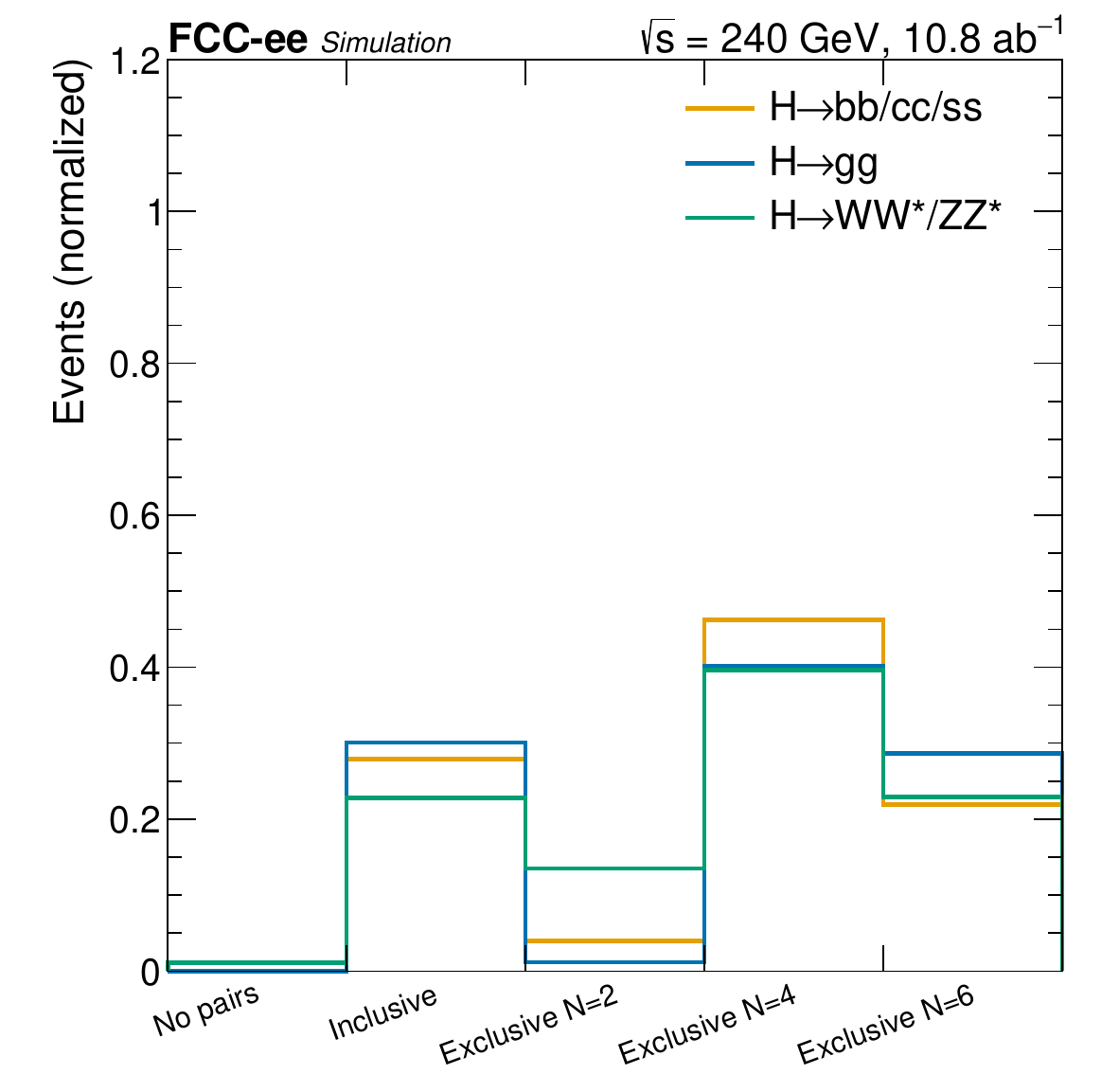}
\includegraphics[width=0.325\linewidth]{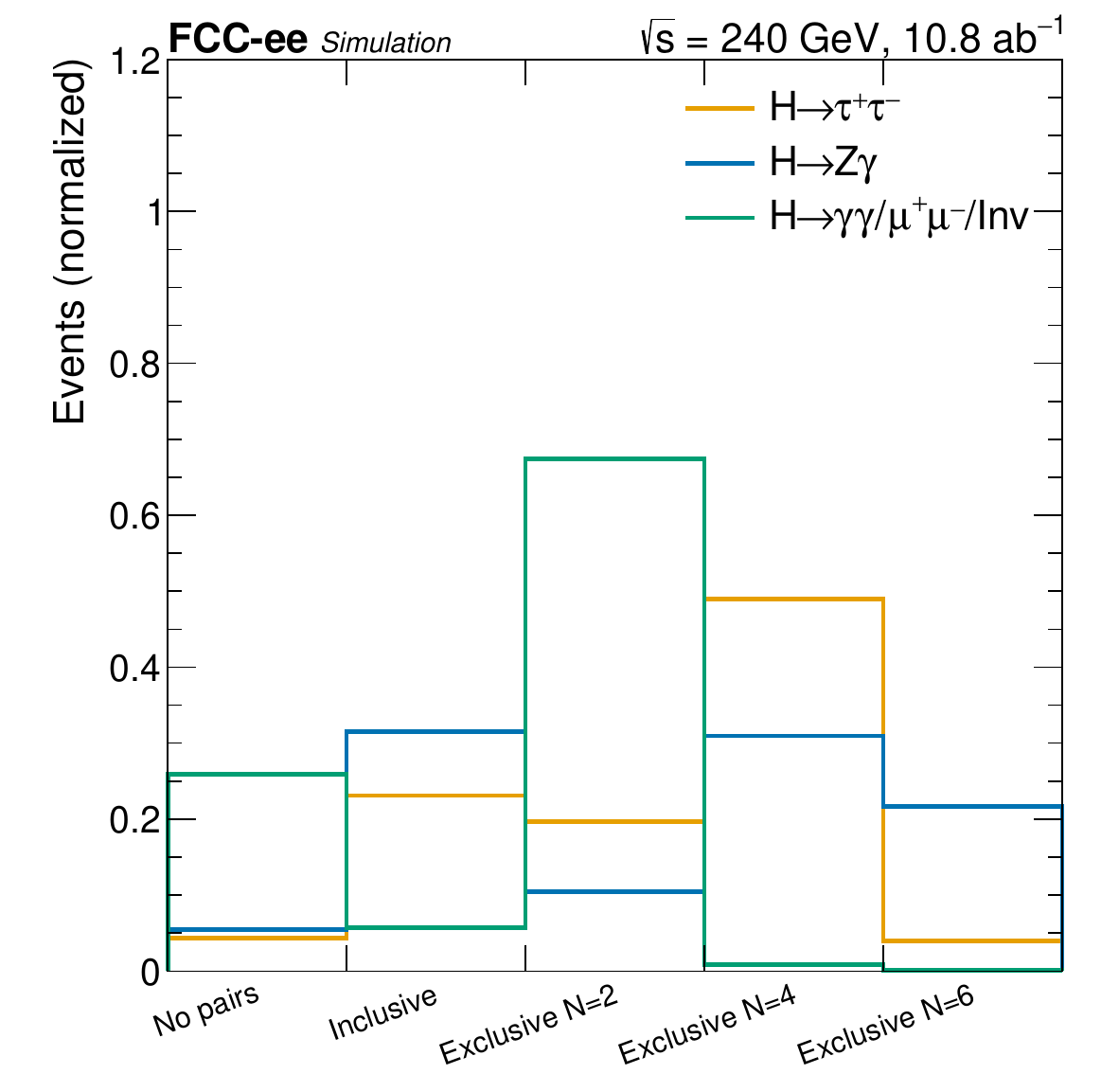}
\includegraphics[width=0.325\linewidth]{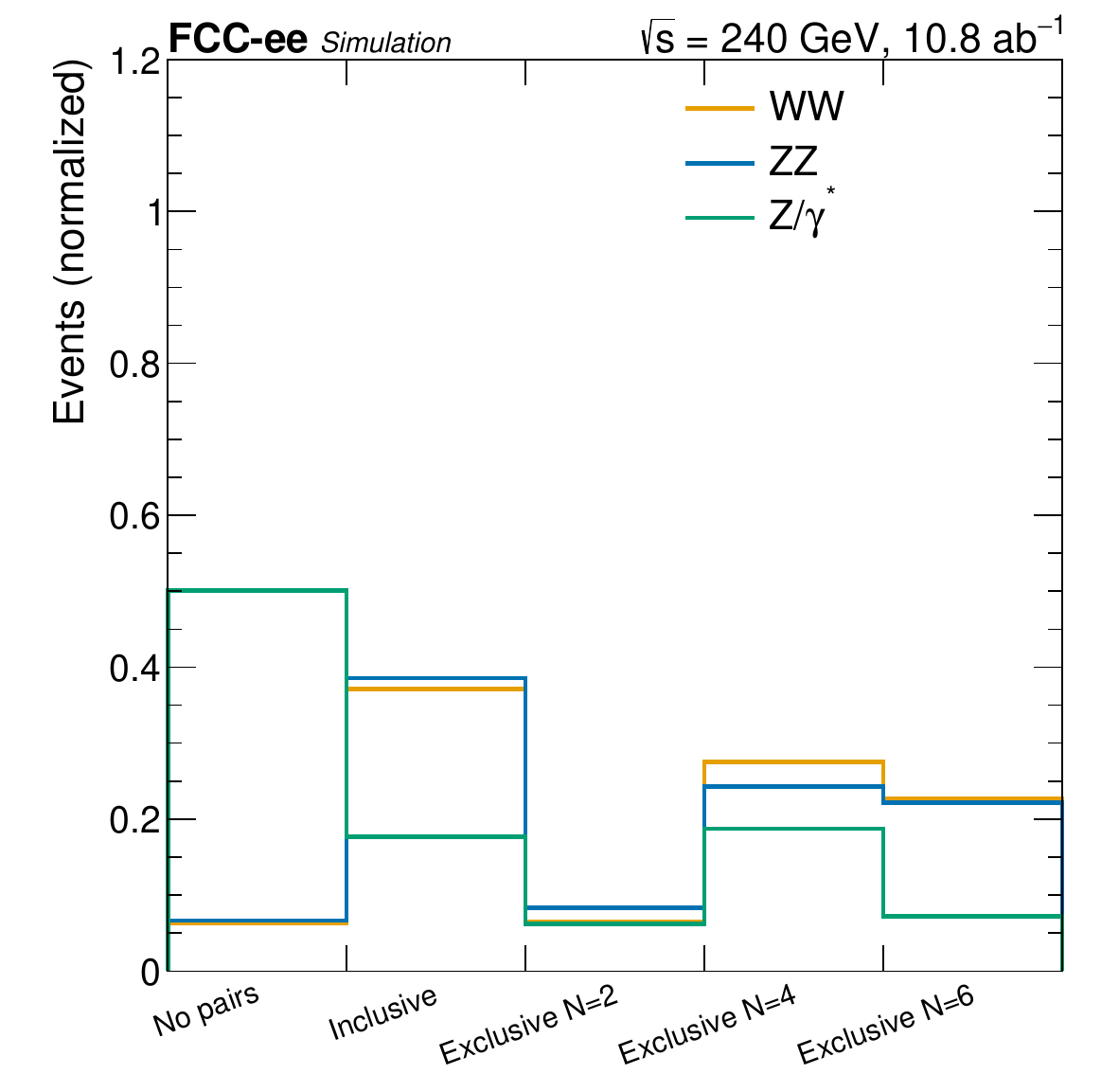}
\includegraphics[width=0.325\linewidth]{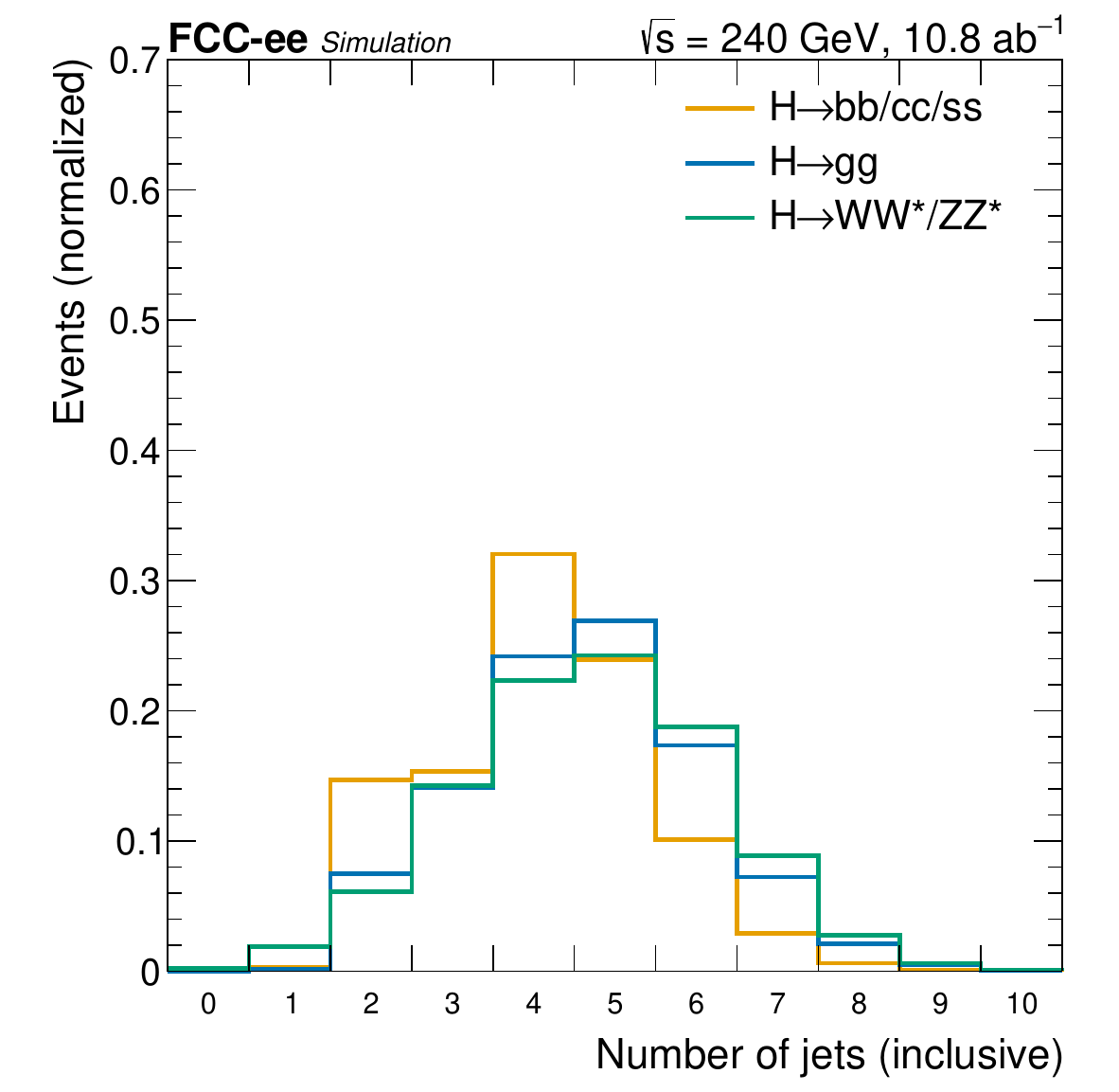}
\includegraphics[width=0.325\linewidth]{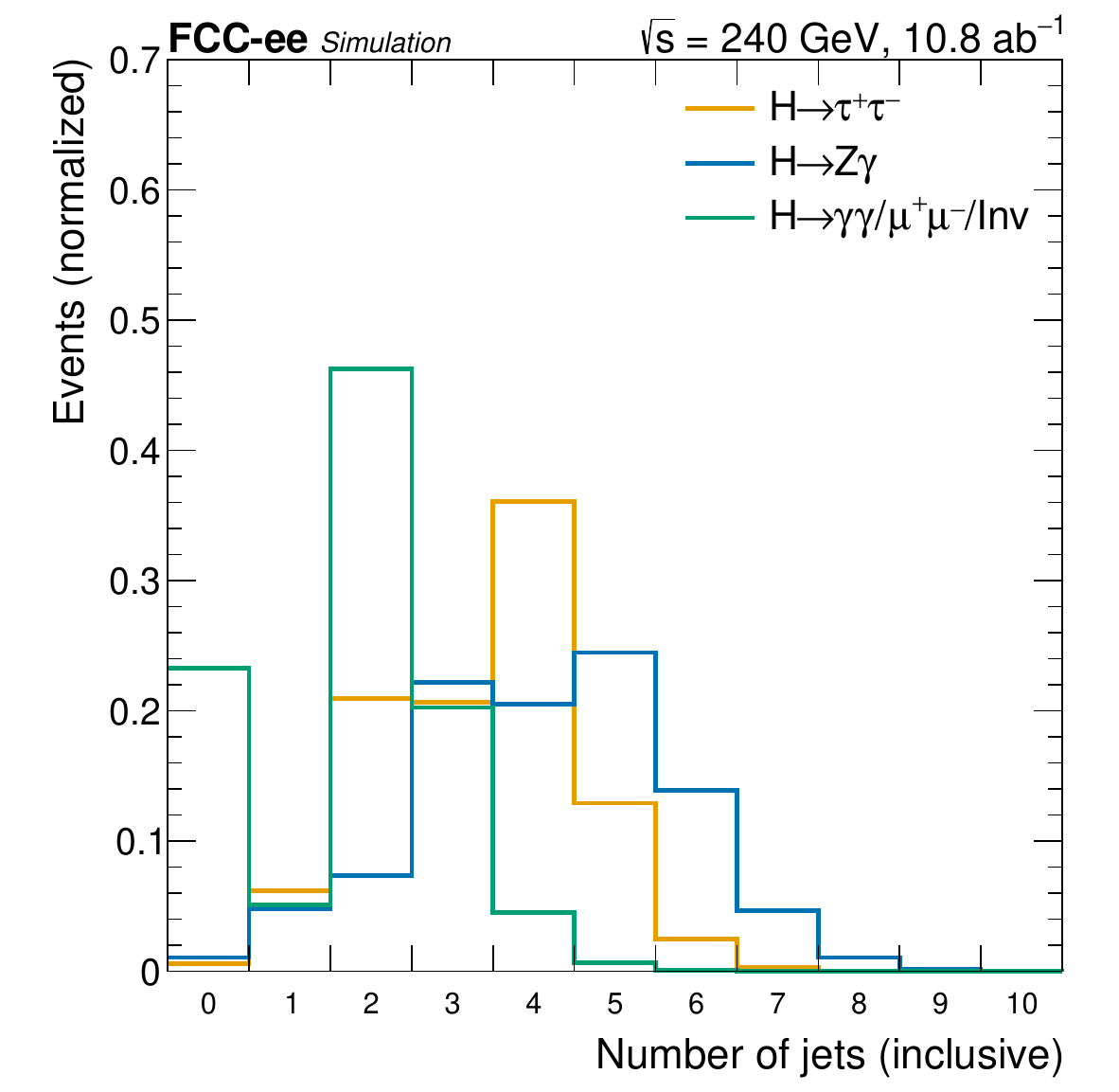}
\includegraphics[width=0.325\linewidth]{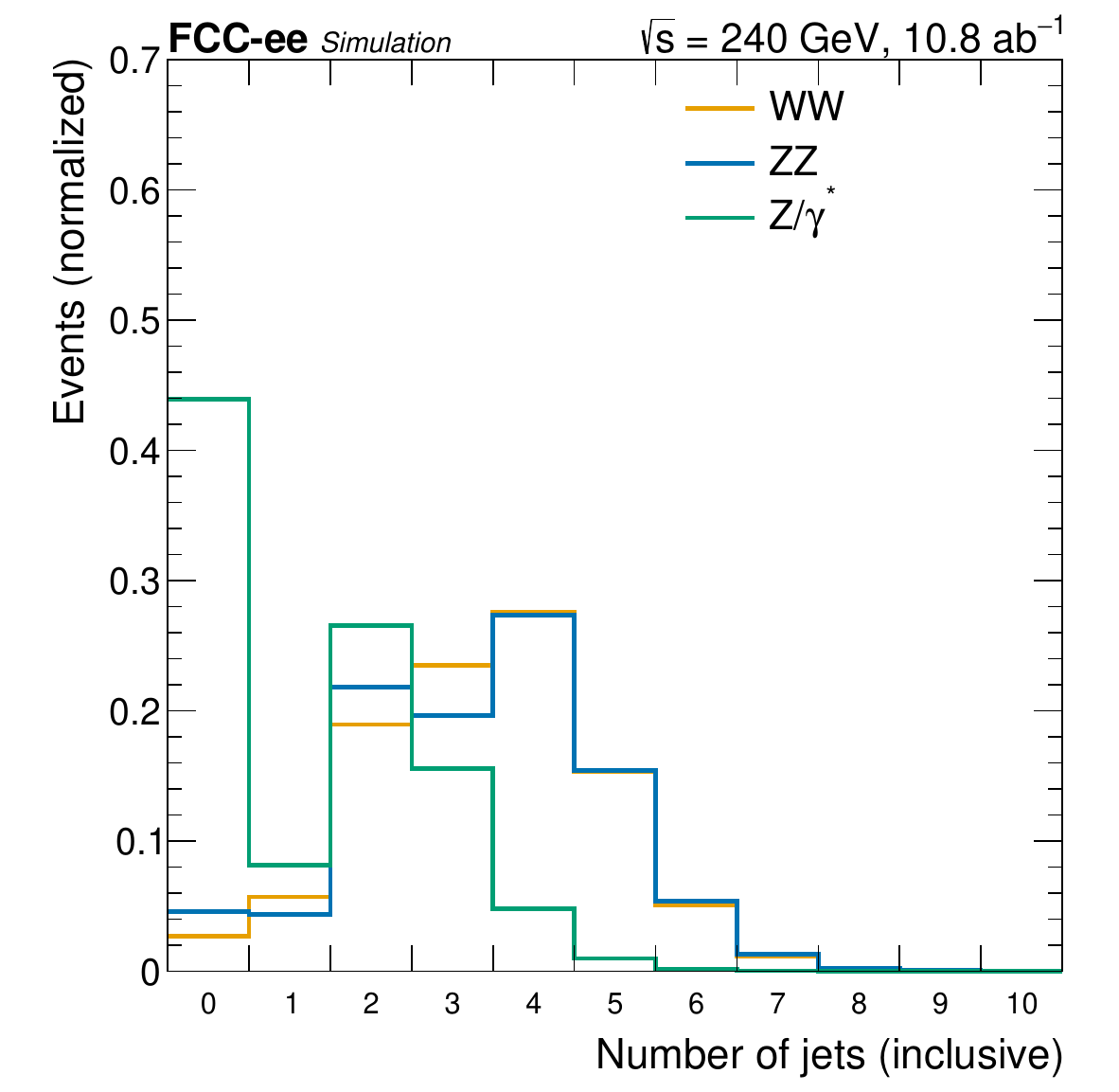}
\caption{Outcome of the jet clustering for events passing the leptonic veto, i.e., at the ``Clustering'' step of Fig.~\ref{fig:xsection:cutflow}, at \sqrtsZH: the clustering scheme retained by the $\chi^{2}$ minimization of Eq.~(\ref{eq:chi2}) (top), and the jet multiplicity obtained with the inclusive $\rm k_{\mathrm{t}}$ algorithm (bottom). Distributions are shown for the \ZH signal, inclusive in the Z boson decay, separated by Higgs boson decay mode (left, center), grouped by decay topologies leading to a comparable clustering outcome, and for the background processes with inclusive boson decays (right). All distributions are normalized to unity. The distributions are not used in the analysis but illustrate the behavior of the jet clustering among the wide range of event topologies.}
\label{fig:clustering}
\end{figure}

Figure~\ref{fig:clustering} illustrates the behavior of this clustering procedure at \sqrtsZH. The distributions are not used as direct inputs to the analysis, but are shown to verify that the clustering strategy identifies the appropriate jet multiplicity over the wide range of topologies of the hadronic final state. The bottom row shows the jet multiplicity for the inclusive signal and background processes, obtained with the inclusive $\rm k_t$ algorithm, which ranges from two jets for the $\mathrm{H} \to \gamma\gamma/\mu\mu/\mathrm{inv.}$ decay modes to four for $\mathrm{H} \to \mathrm{b\bar{b}/c\bar{c}/s\bar{s}/gg}$ and up to six for $\mathrm{H} \to \mathrm{WW^*/ZZ^*}$, while the diboson backgrounds are concentrated at two to four jets. Signal events in which the Z boson decays to neutrinos or leptons have fewer jets, down to none for the $\mathrm{H} \to \gamma\gamma/\mu\mu/\mathrm{inv.}$ decay modes. For $\mathrm{H} \to \mathrm{b\bar{b}/c\bar{c}/s\bar{s}}$ and gg these events contain only the Higgs boson jets and are mostly assigned to the inclusive or $\mathrm{N} = 4$ schemes. The top row shows the clustering scheme retained by the $\chi^2$ minimization of Eq.~(\ref{eq:chi2}) for the same decay modes and backgrounds. For the \ZH signal, the expected topology is observed. 
The exclusive $\mathrm{N} = 2$ scheme is retained mainly for the $\mathrm{H} \to \gamma\gamma/\mu\mu/\mathrm{inv.}$ decay modes. For the other decay modes, most events are assigned to the exclusive $\mathrm{N} = 4$ and $\mathrm{N} = 6$ schemes, consistent with the dominant $\mathrm{H} \to \mathrm{b\bar{b}/c\bar{c}/s\bar{s}}$ (four jets) and $\mathrm{H} \to \mathrm{WW}^{*}/\mathrm{ZZ}^{*}$ (four or six jets) decays.
The inclusive scheme is retained in a minority of the events, and rarely for the $\mathrm{H} \to \gamma\gamma/\mu\mu/\mathrm{inv.}$ decay modes. For these decay modes, the events in which the Z boson decays invisibly or leptonically have no valid jet pair.
%The inclusive scheme is retained for a small fraction of events in all decay modes. 
The backgrounds behave differently, with \Zg events frequently left with no jets or no valid jet pair owing to invisible or leptonic decays, while diboson events are most often assigned to the inclusive scheme, followed by the exclusive $\rm N = 4$ and $\rm N = 6$ schemes. The retained scheme follows the inclusive jet multiplicity of the bottom row, confirming that the event-by-event $\chi^2$ minimization selects the appropriate scheme without any decay-mode-specific tuning.

%The result of the jet clustering choice is illustrated in Fig.~\ref{fig:clustering} (top row) for the different Higgs decay modes and the backgrounds. The final category is determined by the $\chi^2$ selection. For the \ZqqH signal, the expected topology is observed: the $\rm N=2$ category is strongly suppressed, except for the $\rm H \to \gamma\gamma/\mu\mu/inv.$ decay modes, while most events populate the exclusive $\rm N=4$ and $\rm N=6$ categories, consistent with the dominant $\rm H \to bb/cc/ss$ (four jets) and $\rm H \to WW^{*}/ZZ^{*}$ (four or six jets) decays of the Higgs boson. The bottom row shows the inclusive jet multiplicity obtained with the inclusive clustering algorithm, which is one of the clustering options considered in the event-by-event procedure. These distributions provide complementary information on the reconstructed jet content for different Higgs decay modes and backgrounds, and help interpret the role of the inclusive clustering option. The inclusive clustering algorithm (Fig.~\ref{fig:clustering}, bottom row) leads to similar qualitative conclusions, with distributions peaking in the range of 2--6 jets depending on the Higgs decay mode. Backgrounds, on the other hand, show broader patterns: \Zg events are often reconstructed with no jets, or no valid jet pairs due to invisible or leptonic decays, while diboson processes predominantly fall into the 2 or 4 jet category.

\subsection{Kinematic selection} \label{para:selection}

Once the leptons or jets from the associated \Z decay have been identified, a kinematic selection is applied to further suppress background contributions and enhance the signal purity. The selection relies exclusively on the kinematic properties of the \Z decay products, thereby minimizing the dependence on the Higgs boson decay mode.

For the leptonic final states, the mass of the lepton pair is required to be consistent with the \Z boson mass: $86 < \mll < 96~\GeV$. The lepton-pair momentum is then constrained to $20 < \pll < 70~\GeV$. At \sqrtsTop, this requirement is adjusted to $50 < \pll < 150~\GeV$ to account for the larger boost of the \Z boson.

For the hadronic final state, a looser selection is applied than in the leptonic case, reflecting the poorer jet energy resolution. At \sqrtsZH (\SI{365}{\GeV}), the jet-pair mass is required to satisfy $20(60) < \mjj < 140(200)$\GeV and $\rm 20 < \pjj < 90(160)$\GeV. The polar angle of the jet-pair system must satisfy $\rm \lvert \cos\theta_{jj} \rvert < 0.85$, and the acollinearity between the two jets, defined as $\rm \Delta\theta_{jj}$, where $\rm \theta_{jj}$ is the opening angle between the two jet directions in the polar plane, must be larger than 0.35 radians. To suppress the abundant WW background, the event is re-clustered with the exclusive Durham algorithm into four jets, which are grouped into two pairs under a W-boson hypothesis: among the three possible pairings, the one minimizing $\rm \chi^2_{\mathrm{W}} = (m_{jj} - m_{\mathrm{W}})^2 + (m'_{jj} - m_{\mathrm{W}})^2$ is chosen, where $\rm m_{jj}$ and $\rm m'_{jj}$ are the two jet-pair masses. Events are required to satisfy $\chi_{\mathrm{W}} > \SI{6}{\GeV}$ for the selected pairing, excluding events in which both jet-pair masses lie within a circle of radius \SI{6}{\GeV} around $\rm m_{\mathrm{W}} = \SI{80.4}{\GeV}$ in the $\rm (m_{jj}, m'_{jj})$ plane. The mass distribution of the pair containing the leading jet is shown in Fig.~\ref{fig:mva:hadronic} (top left), clearly illustrating the compatibility of the \WW background with the W-boson mass. Further rejection of the \Zg background is achieved by requiring $\lvert \cos\theta_{\rm miss} \rvert < 0.995$, where $\theta_{\rm miss}$ is the polar angle of the missing momentum vector. The missing momentum is reconstructed by summing the four-momenta of all detected particles in the event and applying overall energy–momentum conservation. It represents the momentum carried by undetected particles, most notably neutrinos, or by particles that travel along the beam pipe and thus escape detection. In particular, \Zg events are characterized by a hard initial-state radiation photon emitted collinearly with the beam direction. At \sqrtsTop, an additional cut on the event thrust is applied, $\rm T < 0.85$, to suppress \Zg events, where $\text{T} = \max_{\vec{n}} \frac{\sum_i |\vec{p}_i \cdot \vec{n}|}{\sum_i |\vec{p}_i|}$ with the sum running over all particle momenta~$\vec{\mathrm{p}}_i$. For illustration, Fig.~\ref{fig:mva:hadronic} (top right) shows the thrust magnitude distribution at \sqrtsZH, while at \sqrtsTop the thrust for \Zg events peaks at higher values due to the larger boost, making a cut more effective.

\begin{figure}[!ht]
\centering
\includegraphics[width=0.49\linewidth]{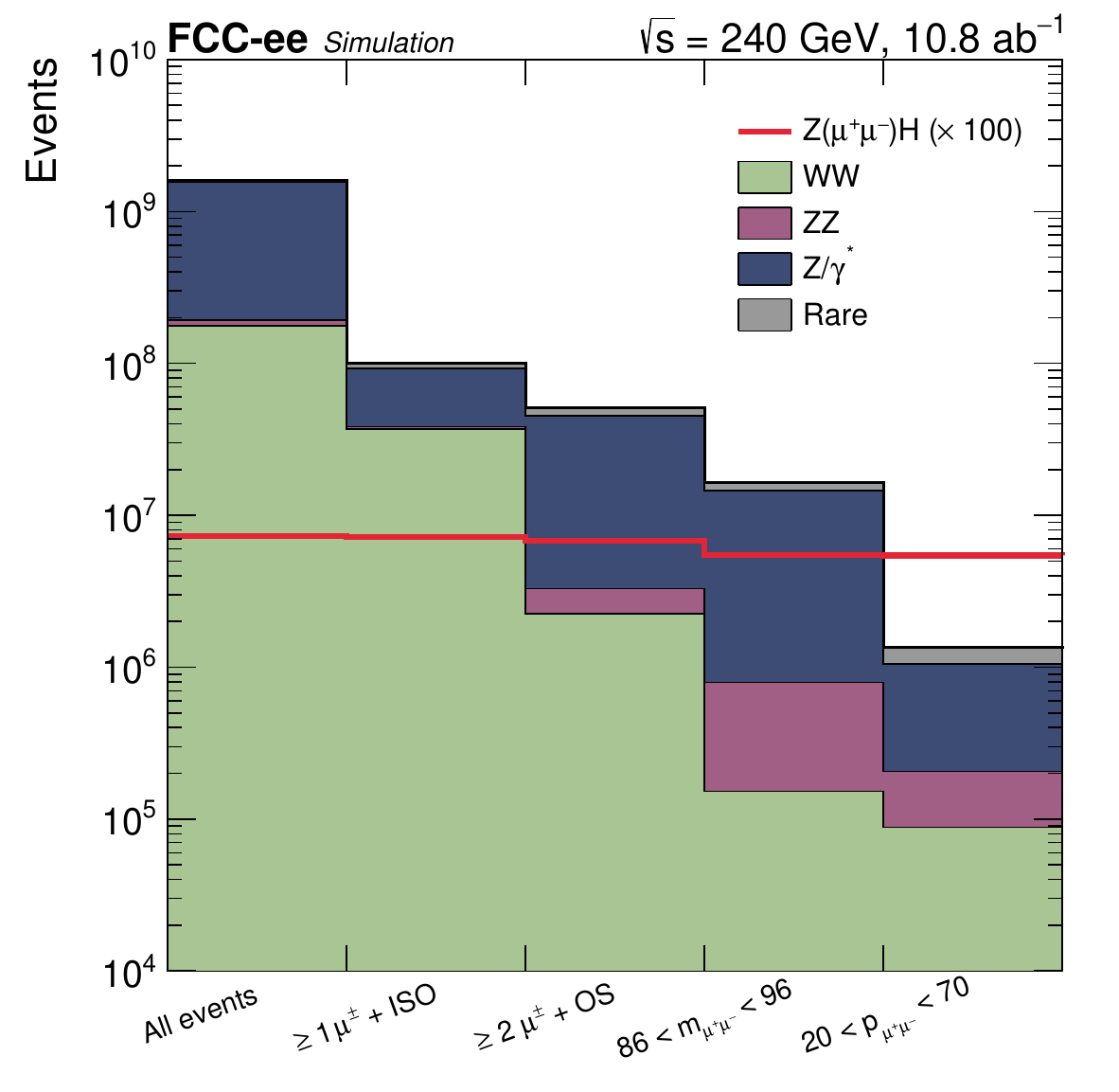}
\includegraphics[width=0.49\linewidth]{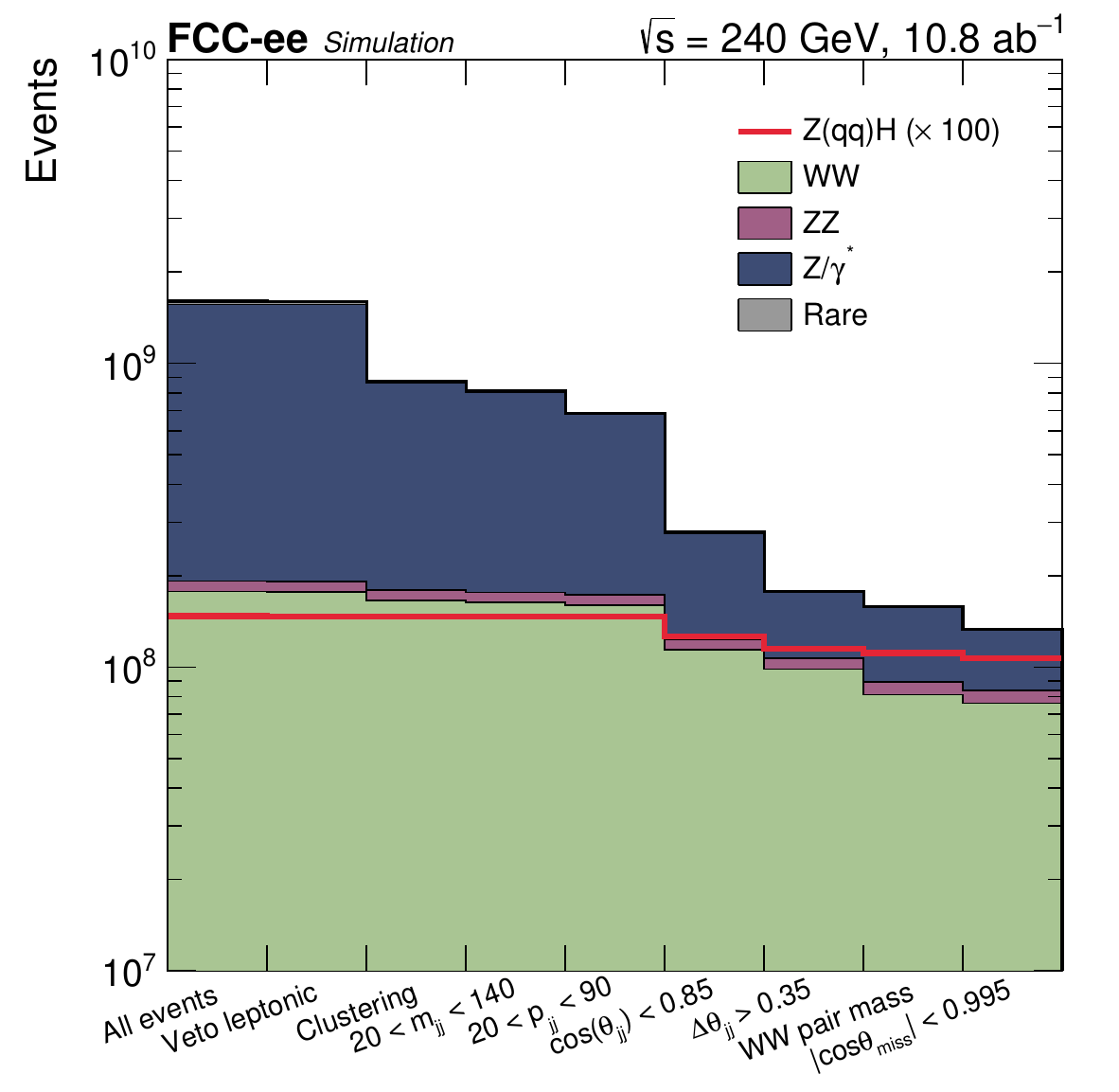}
\caption{Event selection cut flow at \sqrtsZH for the muon (left) and hadronic (right) final states. The signal process corresponds to the \ZmumuH and \ZqqH processes, respectively.}
\label{fig:xsection:cutflow}
\end{figure}

\begin{figure}[!ht]
\centering
\includegraphics[width=0.49\linewidth]{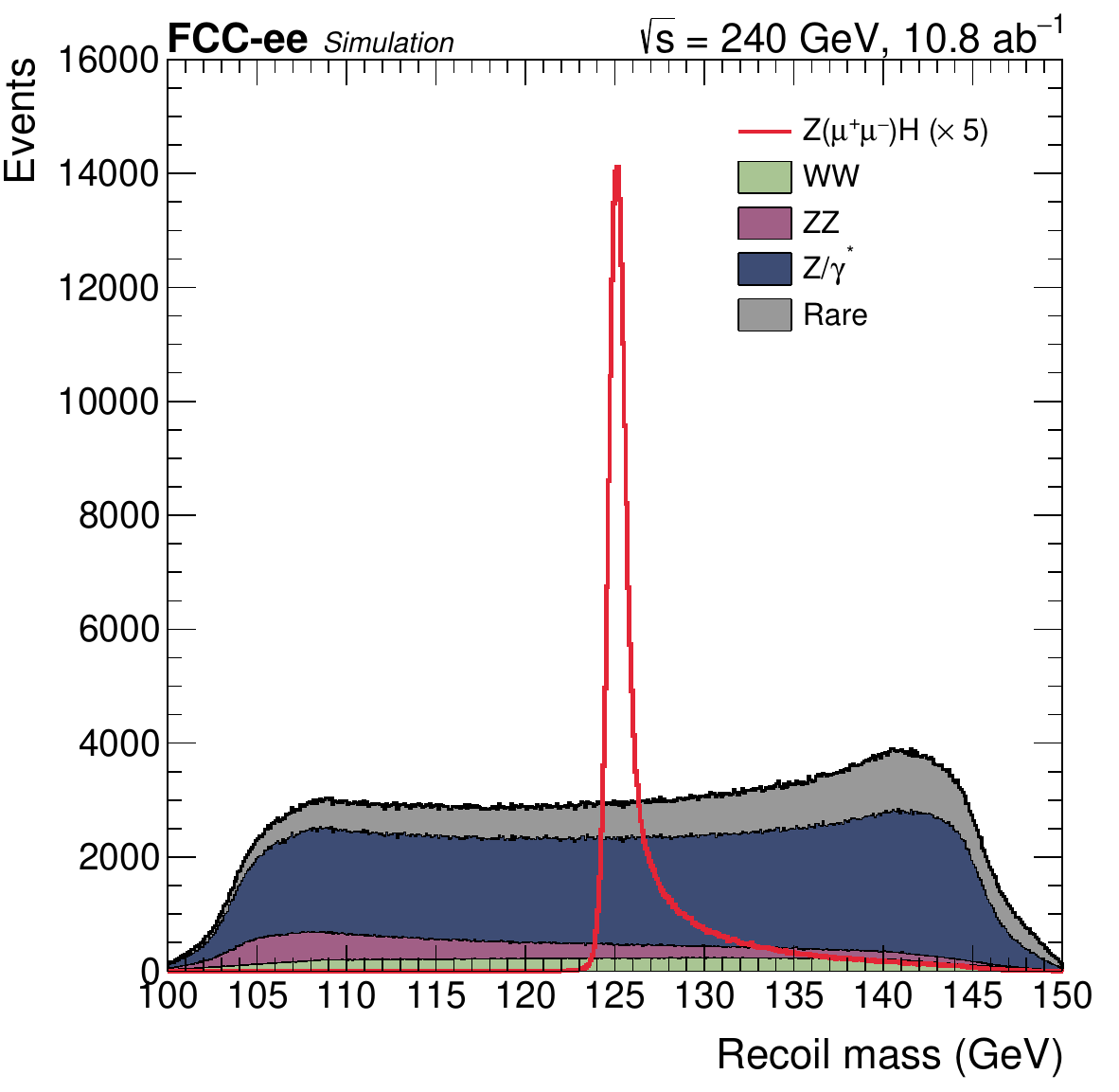}
\includegraphics[width=0.49\linewidth]{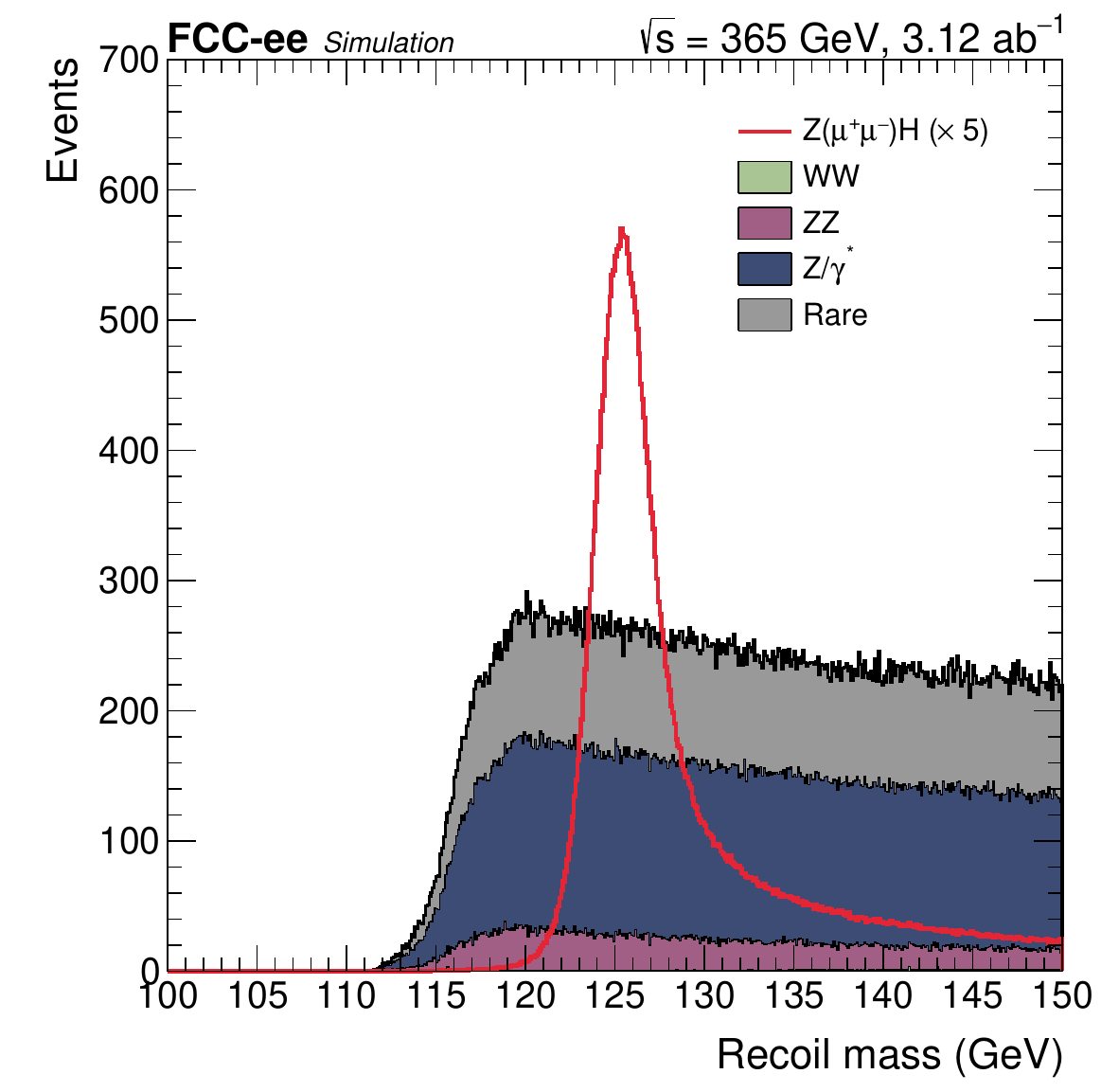}
\caption{Recoil mass distributions for the muon final state at \sqrtsZH (left) and \sqrtsTop (right), after the baseline selections, showing both signal and background contributions.}
\label{fig:xsection:mrecoil}
\end{figure}

The complete event-selection flow, including both object-level and kinematic requirements, is shown in Fig.~\ref{fig:xsection:cutflow} for the muon and hadronic final states at \sqrtsZH. The electron channel exhibits similar behavior and is therefore not shown for brevity. The recoil mass distributions for the muon final state are displayed in Fig.~\ref{fig:xsection:mrecoil} for both \sqrtsZHTop. At \sqrtsTop, the increased center-of-mass energy, larger beam-energy spread, and higher lepton momenta collectively broaden the recoil mass distribution and generate a longer tail toward higher recoil mass values compared to \sqrtsZH.

\subsection{Selection efficiency}\label{para:seleff} 
Studying the signal selection efficiency allows us to assess whether it is largely independent of the Higgs decay mode, which would indicate that different Higgs decays populate the final-state phase space in a similar manner.
If the efficiencies are consistent across all (known) Higgs decays within the quoted measurement uncertainty, the result can be regarded as largely model independent; otherwise, decay-mode–dependent effects may introduce potential biases. A comprehensive assessment of the model independence of the analysis is performed through a series of statistical bias tests against all known Higgs decay modes, taking into account all the \ZH processes, as discussed in Section~\ref{sec:Model-independent}.

\begin{figure}[!ht]
\centering
\includegraphics[width=0.49\linewidth]{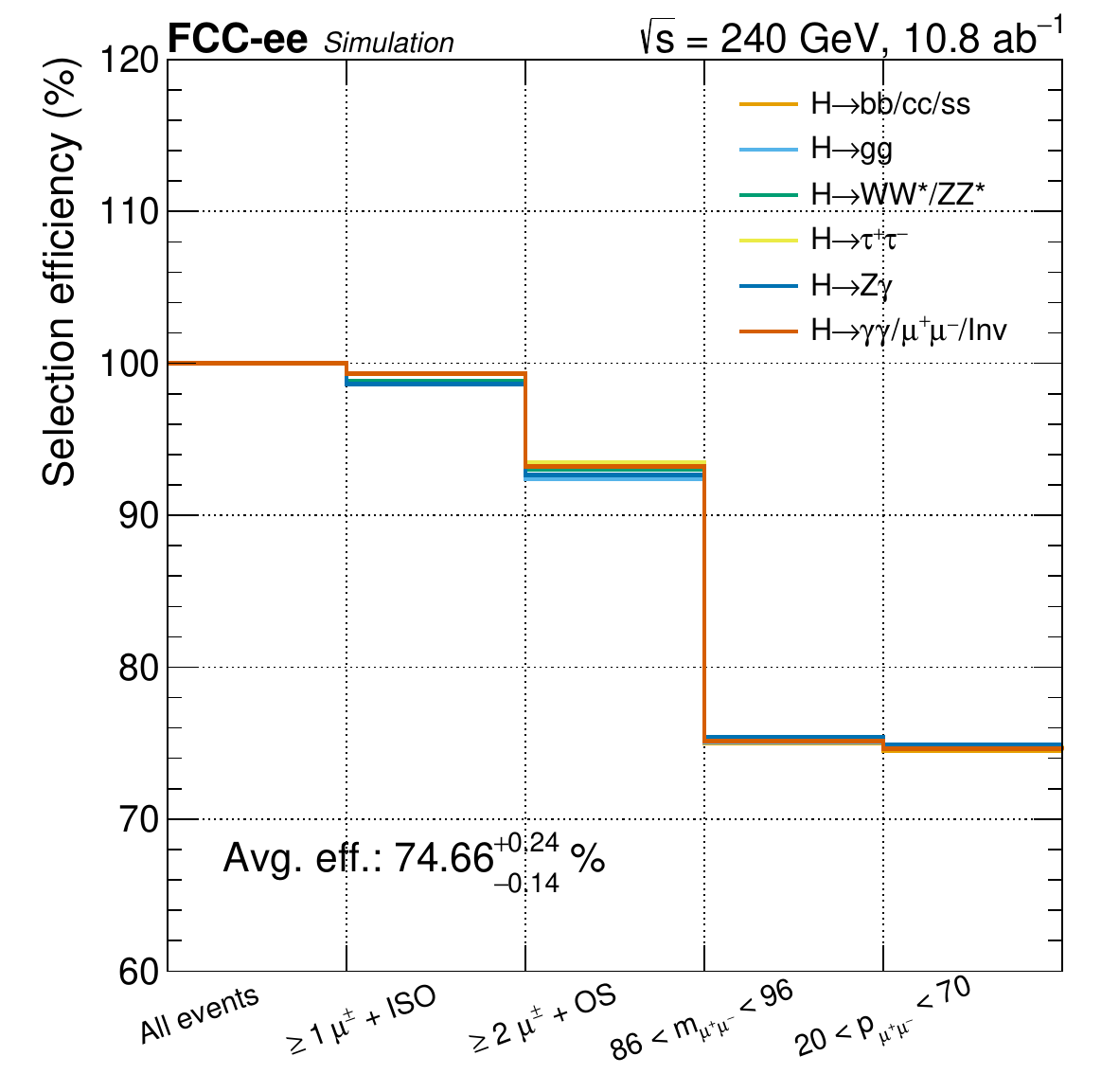}
\includegraphics[width=0.49\linewidth]{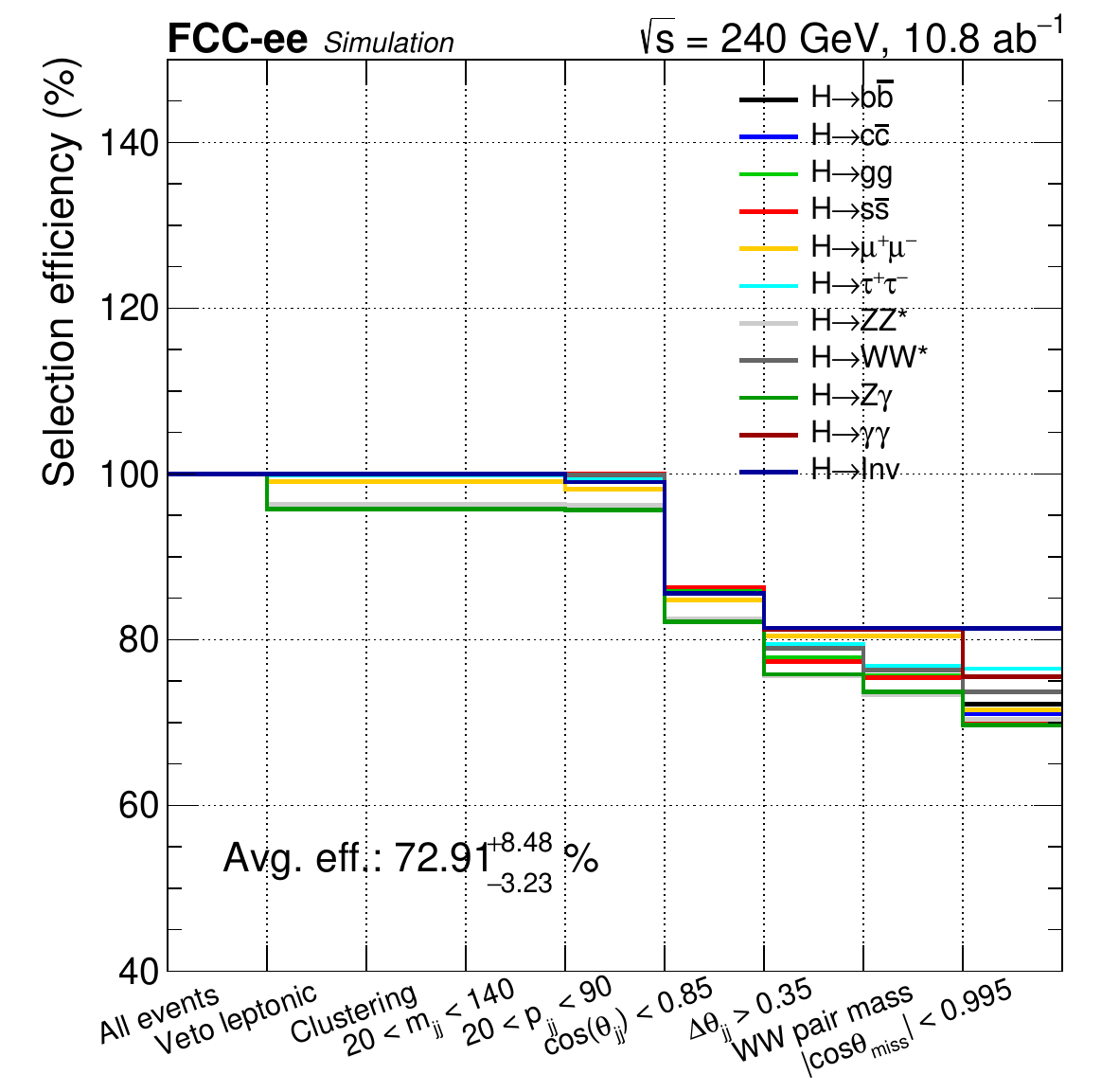}
\caption{Selection efficiencies for different Higgs decay modes at \sqrtsZH for the muon (left) and hadronic (right) final states. The signal corresponds to \ZmumuH and \ZqqH, respectively.}
\label{fig:sel_eff}
\end{figure}

The selection efficiencies for various Standard Model Higgs decay modes are shown in Fig.~\ref{fig:sel_eff} for the muon and hadronic final states at \sqrtsZH, grouped by channels with similar event topologies. In the muon final state, the selection efficiency is nearly identical across all Higgs decay modes at each step of the selection, leading to a final efficiency of 74.66\% with a spread of ${}^{+0.24}_{-0.14}$\% as shown in Fig.~\ref{fig:seleff}. A similar behavior is observed in the electron final state, with a final efficiency of 66.08\% and a spread of ${}^{+0.26}_{-0.29}$\%. The lower efficiency arises from the worse momentum resolution of electrons compared to muons, which broadens the distributions and rejects more events. 
%The lower efficiency arises partly from the ZZ-fusion contribution to the $\mathrm{e^+e^-H}$ final state, which rarely passes the selection, and partly from the worse momentum resolution of electrons compared to muons, which broadens the distributions and rejects more events.
In both cases, the small spread can be interpreted as the residual level of dependence, setting the precision at which the analysis remains effectively independent of the Higgs decay mode.  

In contrast, the hadronic final state exhibits a larger decay-mode dependence. Already at the leptonic veto stage, the efficiency for \HZa and \HZZ decay modes decreases by a few percent due to leptonic \Z decays. This effect is reduced for \Hmumu by explicitly vetoing such events from the leptonic final states while retaining them in the hadronic final state if the dilepton mass is close to \mh (within 3 GeV, see Section~\ref{para:objects}). Larger variations arise from selections on angular distributions and the polar angle of the missing momentum. For different Higgs decay modes, the average final efficiency is 72.91\%, with a spread of ${}^{+8.48}_{-3.23}$\% between the Higgs decay modes as shown in Fig.~\ref{fig:seleff}. This spread exceeds the requirement for decay-mode–independent selection, as expected due to the overlap between the associated hadronic \Z final state and the predominantly hadronic Higgs decay products. Nevertheless, the selection is designed to minimize this dependence as much as possible. The impact of decay-mode dependence is quantified through dedicated bias tests (see Section~\ref{sec:Model-independent}).

The migration between the final states is summarized in Table~\ref{tab:migration}, which gives the efficiency for the three targeted Z boson decay modes of the signal to be selected in the electron, muon, and hadronic final states after the full event selection. Migration between the electron and muon final states, and from Z($\mathrm{q\bar{q}}$)H into the leptonic final states through Higgs boson decays with leptons, is below 0.1\%. In contrast, 25\% of Z($\mathrm{e^+e^-}$)H and 19\% of Z($\mu^+\mu^-$)H events are selected in the hadronic final state, since the hadronic selection is the complement of the leptonic one followed by loose requirements on the jet kinematics only. These events are part of the signal template of the hadronic fit and of the
model-independence tests of Section~\ref{sec:Model-independent}.

\begin{table}[ht!]
\centering
\small
\renewcommand{\arraystretch}{1.1}
\caption{Selection efficiencies (in \%) after the full event selection at \sqrtsZH for the three targeted Z boson decay modes of the ZH signal in the electron, muon, and hadronic final states. The diagonal entries are the efficiencies of the targeted final states quoted in the text, and the off-diagonal entries quantify the migration between final states.}
\vspace{2mm}
\label{tab:migration}
\begin{tabular}{lccc}
\hline
 & \multicolumn{3}{c}{\textbf{Selected final state}} \\
\cline{2-4}
\textbf{Signal process} & \textbf{Electron} & \textbf{Muon} & \textbf{Hadronic} \\
\hline
Z($\mathrm{e^+e^-}$)H   & 66.08 & 0.08  & 25.11 \\
Z($\mu^+\mu^-$)H        & 0.07  & 74.66 & 19.17 \\
Z($\mathrm{q\bar{q}}$)H & 0.07  & 0.08  & 72.91 \\
\hline
\end{tabular}
\end{table}
\section{Signal Extraction}\label{sec:SignalExtraction}

After the baseline event selection that suppresses backgrounds and enhances the signal, the sensitivity is further improved using multivariate techniques and a shape-based fit to signal-sensitive observables.

\subsection{Multivariate analysis}\label{subsec:BDT}

Signal–background separation is enhanced using a Boosted Decision Tree (BDT)~\cite{Roe:2004na} implemented with {\tt XGBoost}~\cite{Chen:2016:XST:2939672.2939785}. Independent Monte Carlo samples of signal and background are used exclusively for training and validation. To minimize bias, the dataset is randomly divided into equal parts for training and validation. The BDT hyperparameters are tuned to optimize performance while preventing overtraining.

\begin{figure}[!ht]
  \centering
  \includegraphics[width=0.49\linewidth]{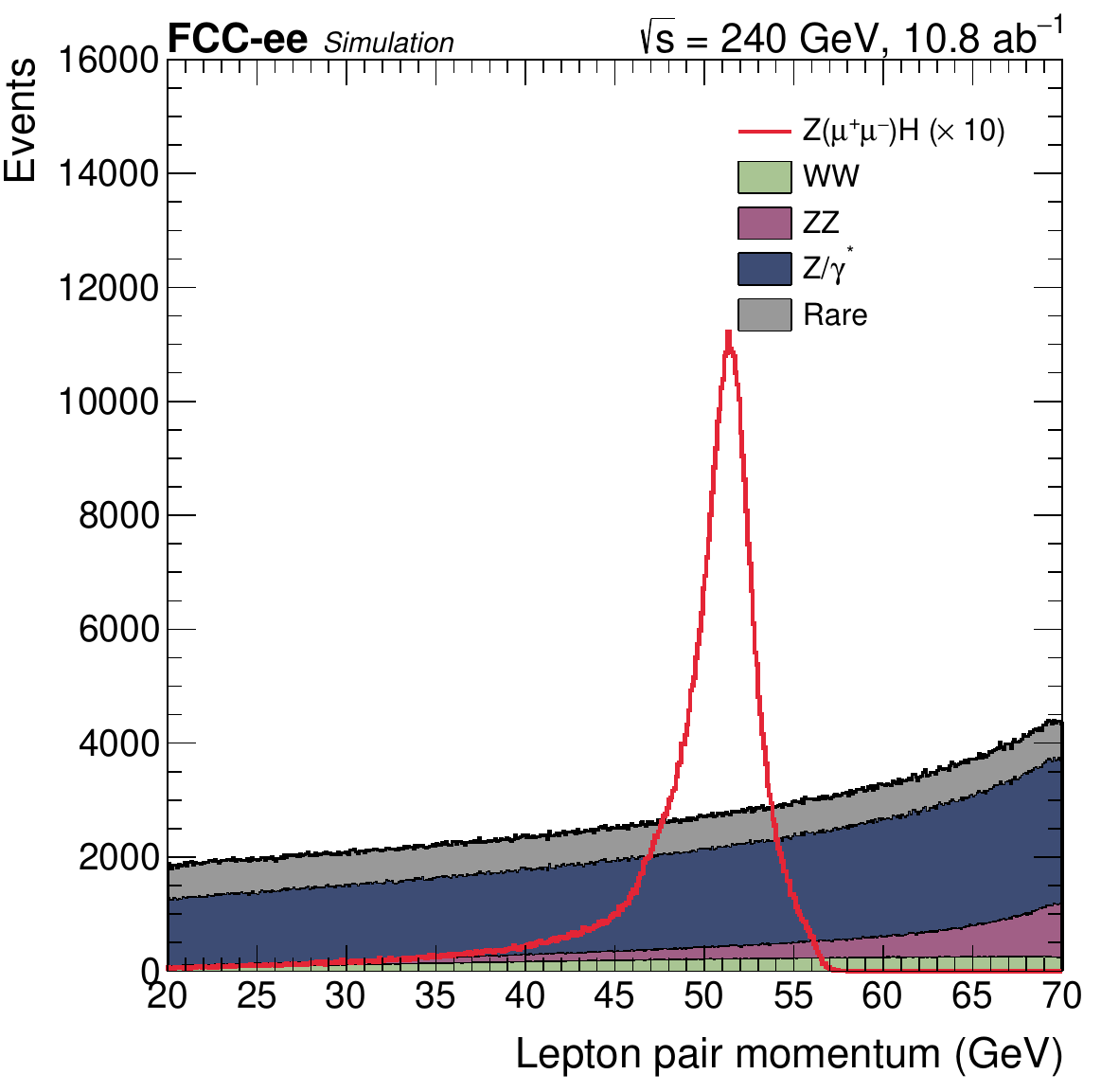}
  \includegraphics[width=0.49\linewidth]{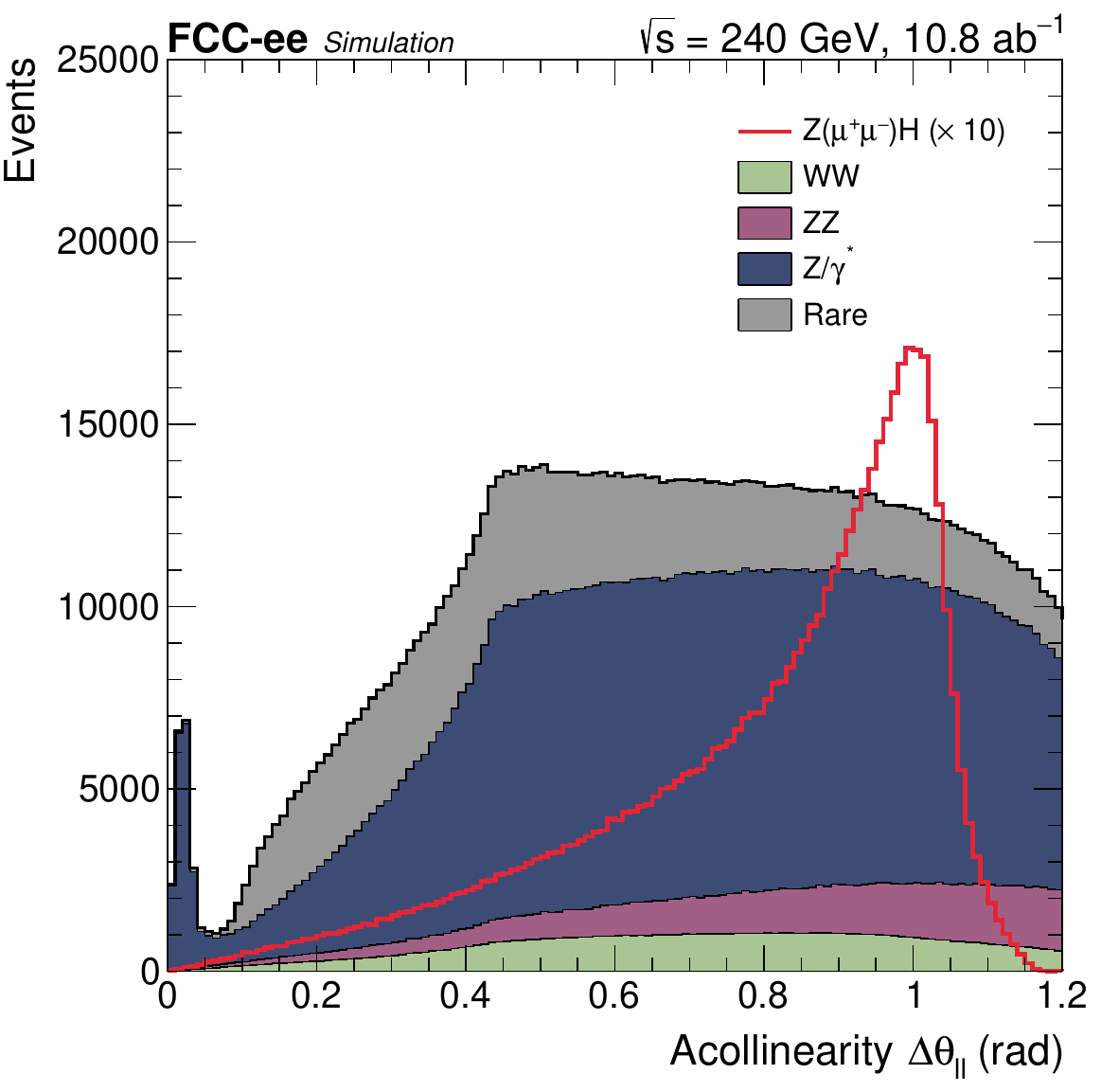}

  \vspace{1em} 
  
  \includegraphics[width=0.49\linewidth]{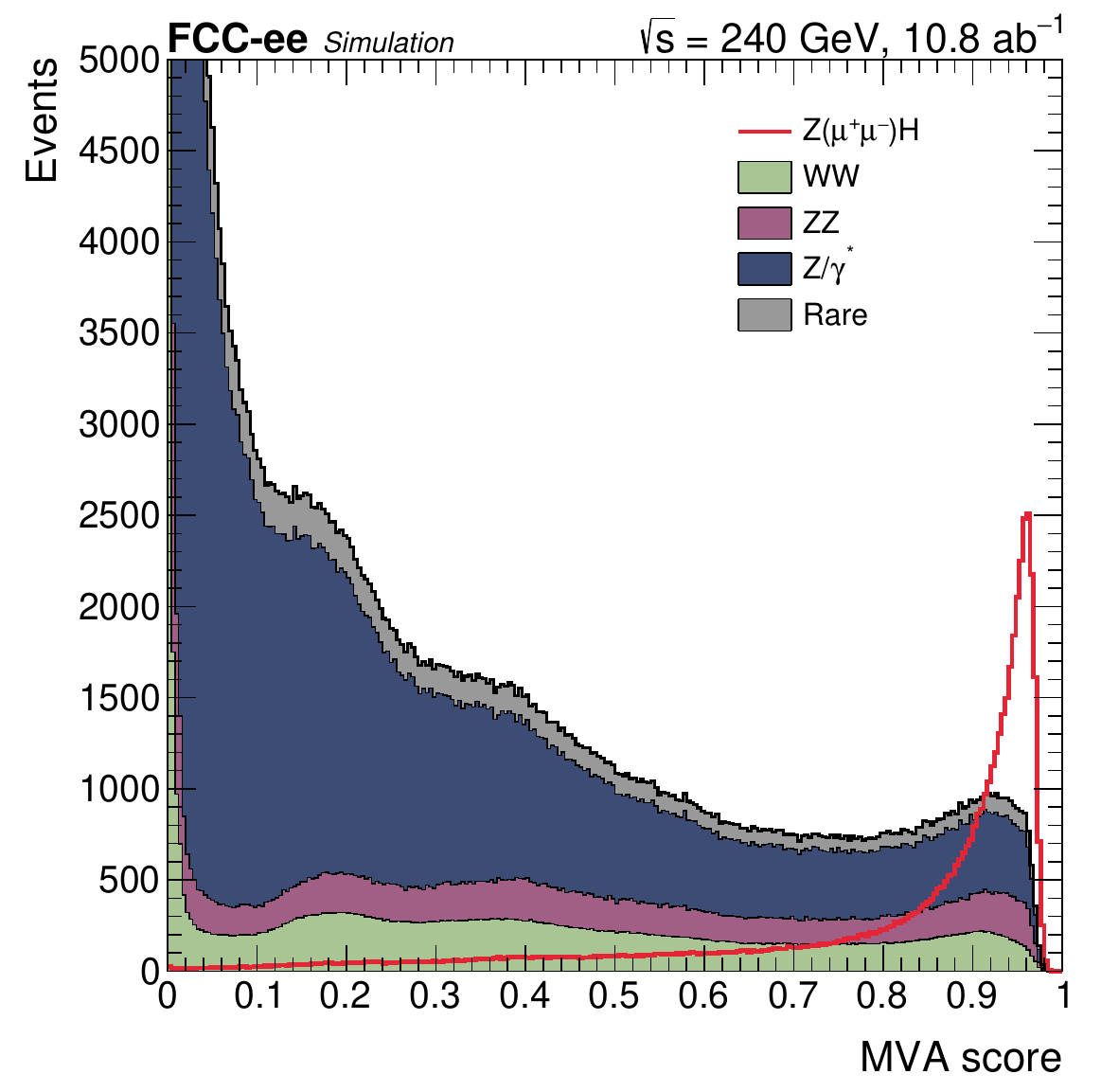}
  \includegraphics[width=0.49\linewidth]{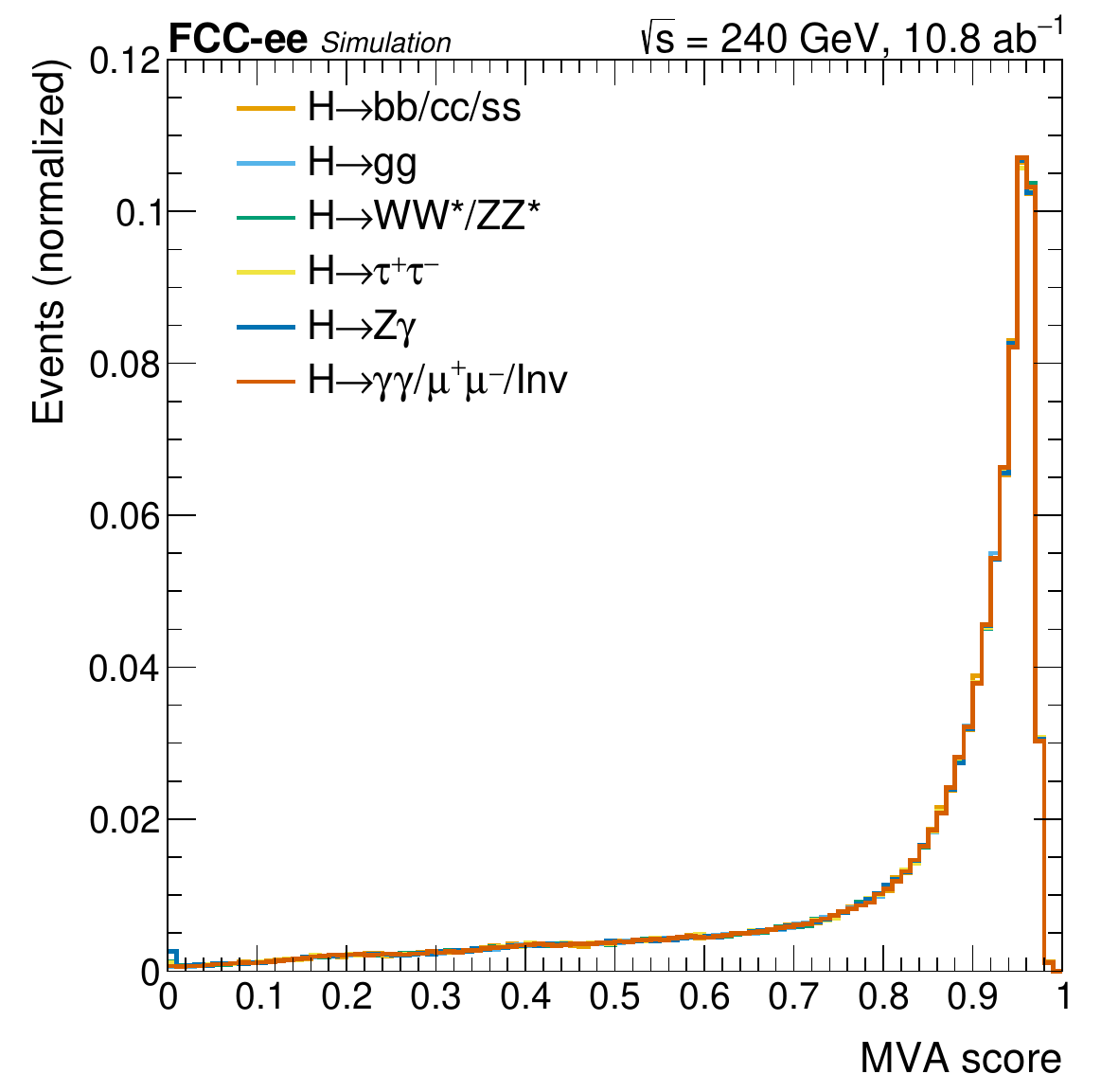}
  \caption{The two leading MVA variables (top row), the final BDT output distribution (bottom left), and the normalized BDT output distributions for the different Higgs decay modes (bottom right) for the muon final state at \sqrtsZH.}
  \label{fig:mva:leptonic}
\end{figure}

In the leptonic analysis, nine input variables are used, all based on the information of the two selected leptons and in order to avoid event-shape variables that could be sensitive to the Higgs boson decay. The inputs include the lepton-pair invariant mass \mll, momentum \pll, and polar angle $\rm \theta_{\ell\ell}$, as well as the momenta and polar angles of the individual leptons. In addition, angular observables such as the acoplanarity ($\rm \pi-\Delta\phi_{\ell\ell}$) and acollinearity ($\rm \Delta\theta_{\ell\ell}$), where $\Delta\phi_{\ell\ell}$ is the difference in azimuthal angle between the two leptons and $\rm \Delta\theta_{\ell\ell}$ is their opening angle in the polar plane, are included. The most discriminating input variables are the momentum of the lepton pair ($\rm p_{\ell\ell}$) and acollinearity (see top row in Fig.~\ref{fig:mva:leptonic} for the muon final state), as well as the polar angle of the leading lepton $\theta_{\ell1}$ and acoplanarity. The resulting BDT output distributions for signal and background are shown in Fig.~\ref{fig:mva:leptonic} (bottom left) for the muon final state at \sqrtsZH, demonstrating clear separation, with an area under the ROC curve (AUC) of about 0.95. Since only kinematic information from the two selected leptons is used, the discriminator consequently minimizes dependence on the Higgs boson decay mode, as illustrated in Fig.~\ref{fig:mva:leptonic} (bottom right). The BDT for the electron final state has a similar performance and features to those in the muon final state.

\begin{figure}[!ht]
  \centering
  \includegraphics[width=0.49\linewidth]{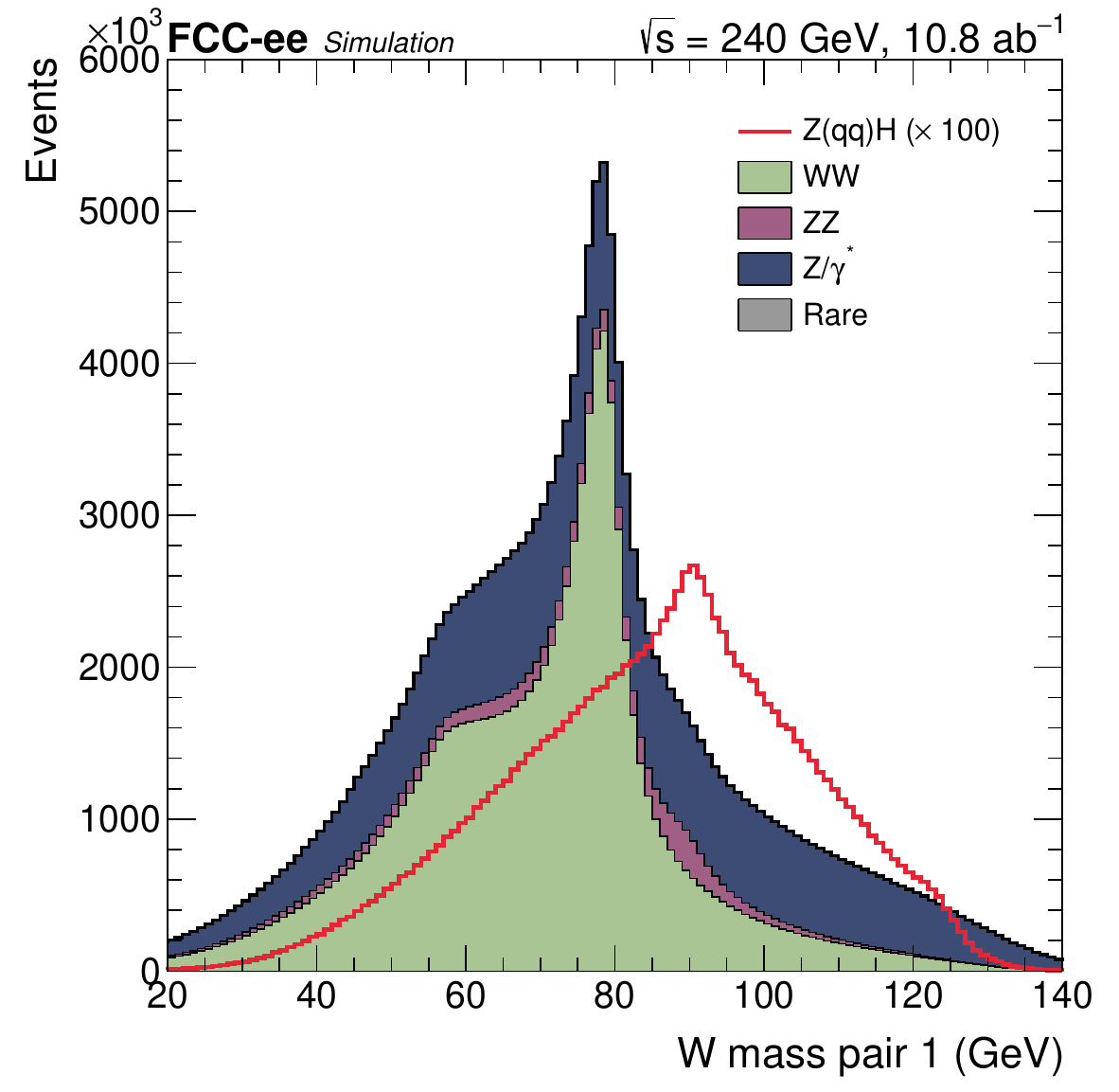}
  \includegraphics[width=0.49\linewidth]{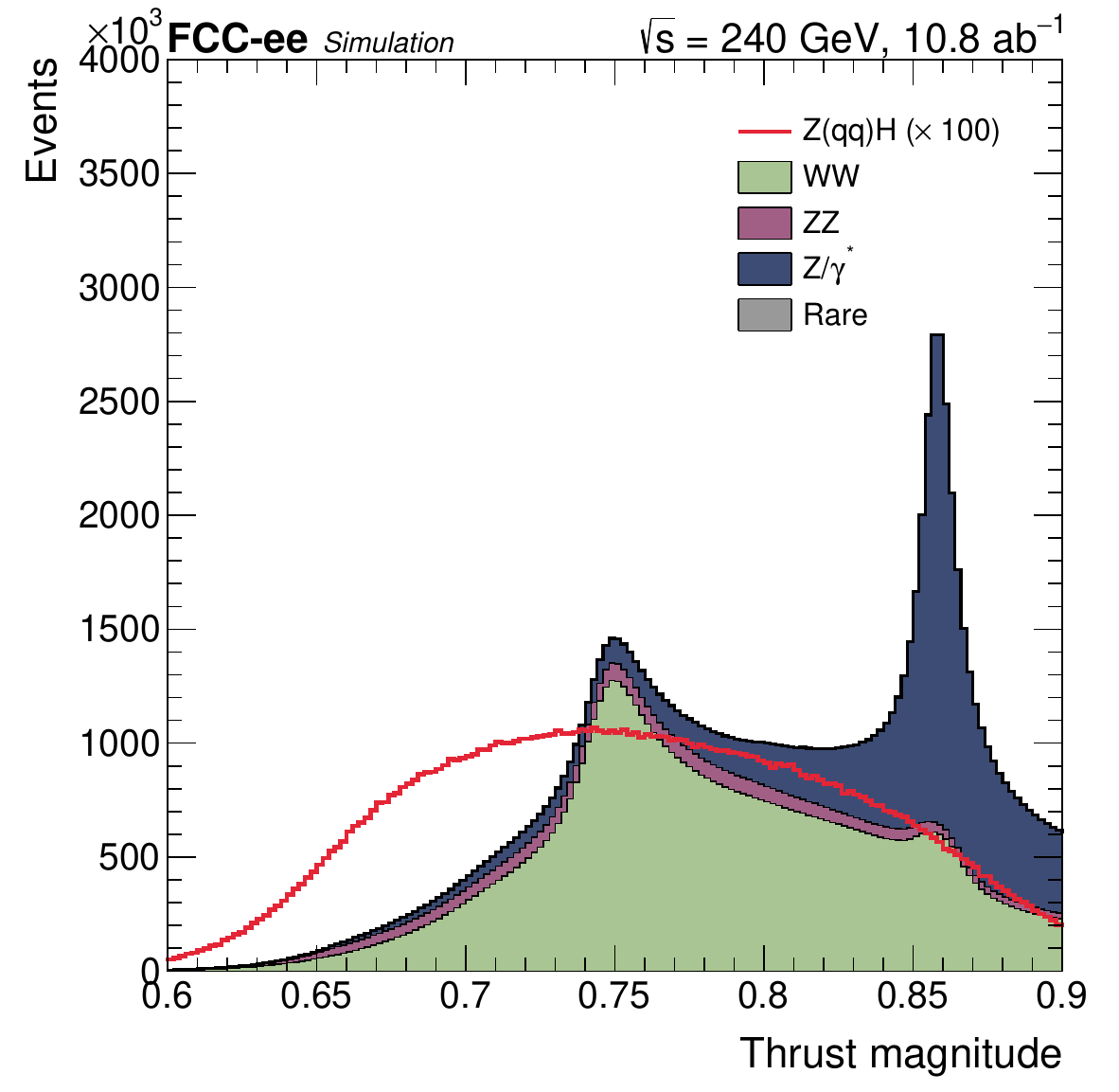}

  \vspace{1em} 

  \includegraphics[width=0.49\linewidth]{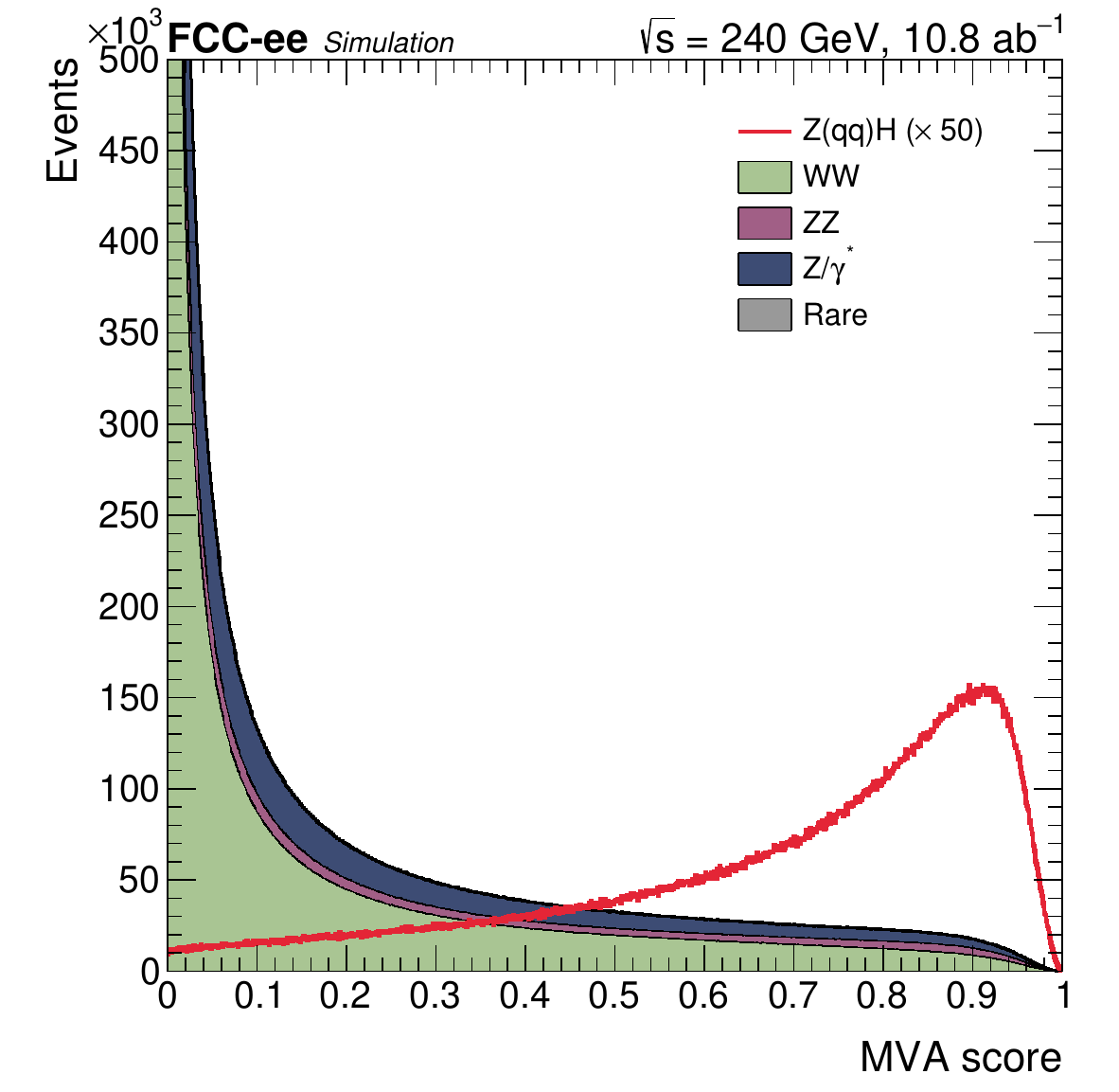}
  \includegraphics[width=0.49\linewidth]{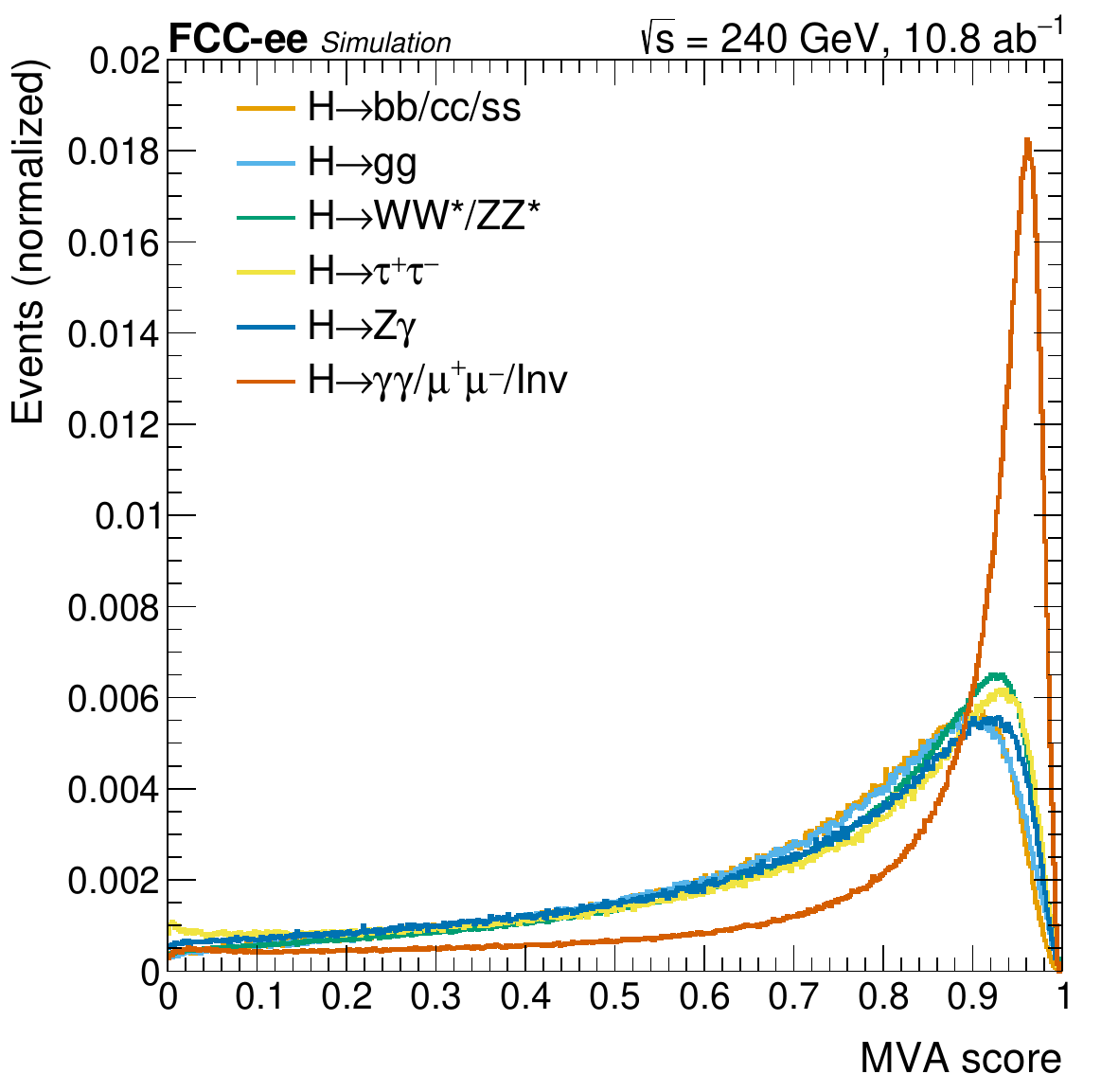}

  \caption{The two leading MVA variables (top row), the final BDT output distribution (bottom left), and the normalized BDT output distributions for the different Higgs decay modes (bottom right) for the hadronic final state at \sqrtsZH. The W mass pair distribution (top left) is shown without the two-dimensional cut on the W mass pairs.}
  \label{fig:mva:hadronic}
\end{figure}

The input variables for the hadronic analysis are largely analogous to those of the leptonic final states. They include the momentum and polar angle ($\rm \cos\theta$) of the jet-pair system, as well as its acoplanarity and acollinearity. The acoplanarity ($\rm \pi - \Delta\phi_{jj}$) and acollinearity ($\rm \Delta\theta_{jj}$) are defined in the same way as in the leptonic case, where $\rm \Delta\phi_{jj}$ is the difference in azimuthal angle between the two jets and $\rm \Delta\theta_{jj}$ is their opening angle in the polar plane. The jet-pair mass is excluded from the training since the hadronic channel contribution is extracted from a fit in the \mjj–\mrec two-dimensional plane, which will be discussed in detail in Section~\ref{subsec:Fit}. At the jet level, the individual jet momenta and polar angles are included. Other variables include the kinematics of the two jet pairs obtained after exclusive re-clustering of the event into four jets, paired under the W-boson hypothesis by the $\rm \chi^{2}_{\mathrm{W}}$ minimization of Section~\ref{para:selection}. The pair containing the leading jet is denoted $\rm \mathrm{W}_{1}$ and the other $\rm \mathrm{W}_{2}$, and their momenta, masses, and polar angles ($\rm p_{\mathrm{W}_{1,2}}$, $\rm m_{\mathrm{W}_{1,2}}$ and $\rm \cos\theta_{\mathrm{W}_{1,2}}$) enter the BDT, together with the magnitude of the event thrust. In total, fifteen variables are included in the BDT, among which the two W-mass variables ($\rm m_{W_{1,2}}$) and the thrust are the most discriminating (see Fig.~\ref{fig:mva:hadronic}, top row). No overtraining is observed.

The resulting BDT output distributions for signal and background are shown in Fig.~\ref{fig:mva:hadronic} (bottom left) at \sqrtsZH, demonstrating clear separation with an AUC of 0.94. Since several input variables have different sensitivities to the Higgs decay topologies, the BDT response is expected to retain some dependence on the Higgs decay mode. This is indeed observed in Fig.~\ref{fig:mva:hadronic} (bottom right), where larger variations appear for \Hmumu, \Haa, and \Hinv decays, which involve objects not used in the analysis (either excluded from the clustering or appearing only as missing energy). Consequently, the topology learned by the BDT, dominated by fully hadronic modes, differs from that of these decay modes. The impact of the observed Higgs-decay dependence of the BDT response is quantified by the statistical bias tests discussed in Section~\ref{sec:Model-independent}.

\subsection{Fitting strategy}\label{subsec:Fit}

A binned maximum-likelihood fit is performed to extract the uncertainty on the total \ZH production cross section. The input templates for the leptonic and hadronic final states are described below. The signal template includes all \ZH final states: the targeted \ZmumuH, \ZeeH, and \ZqqH processes, as well as the contributions from \ZtautauH and \ZnunuH events that can pass the selection criteria.
The \nunuH and \eeH fusion processes are included, but their contribution is negligible owing to their very low selection efficiency.
All \ZH contributions are combined and treated as a single parameter of interest in the fit, corresponding to the total \ZH production rate.

\begin{figure}[!ht]
  \centering
  \includegraphics[width=0.49\linewidth]{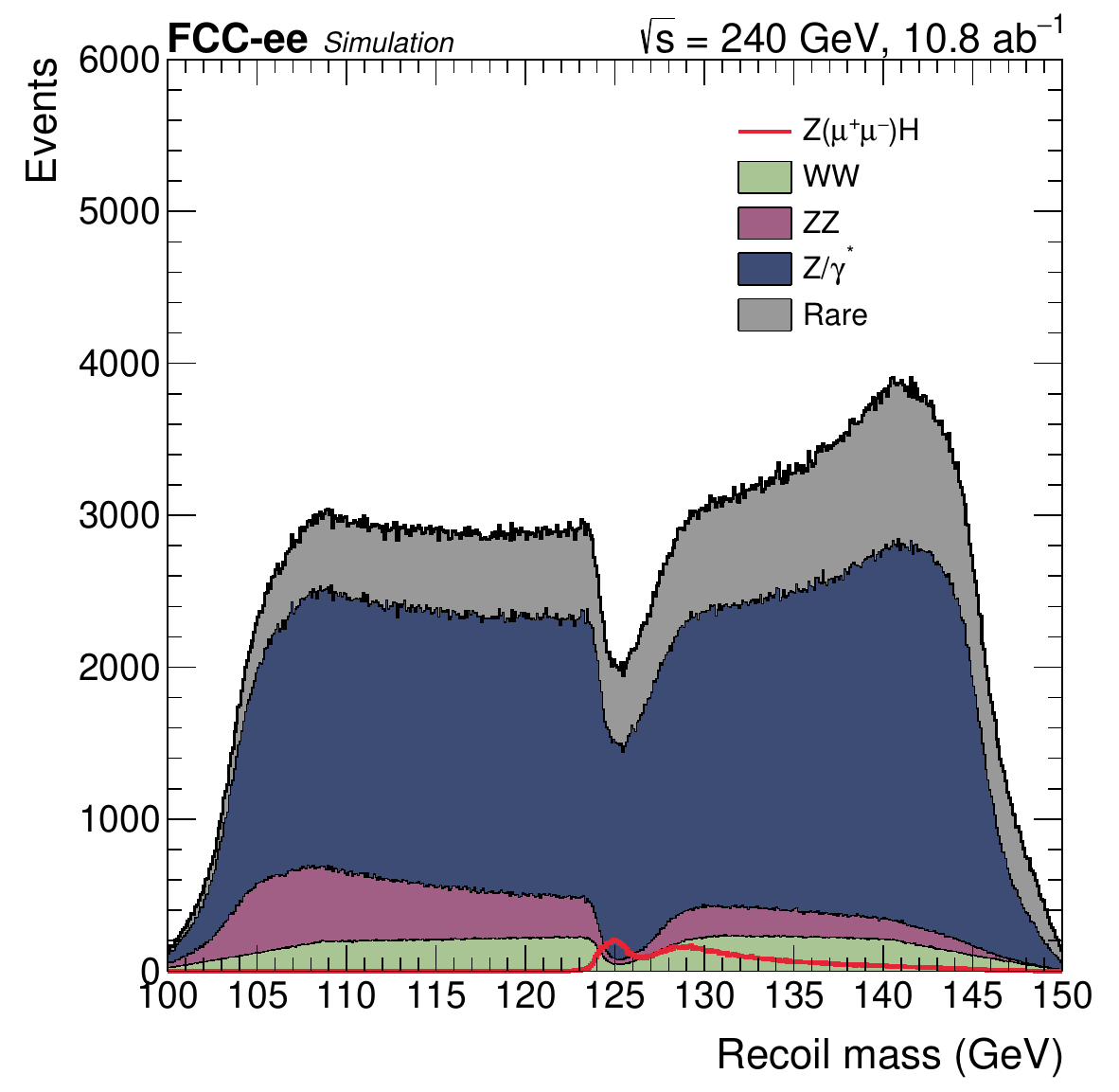}
  \includegraphics[width=0.49\linewidth]{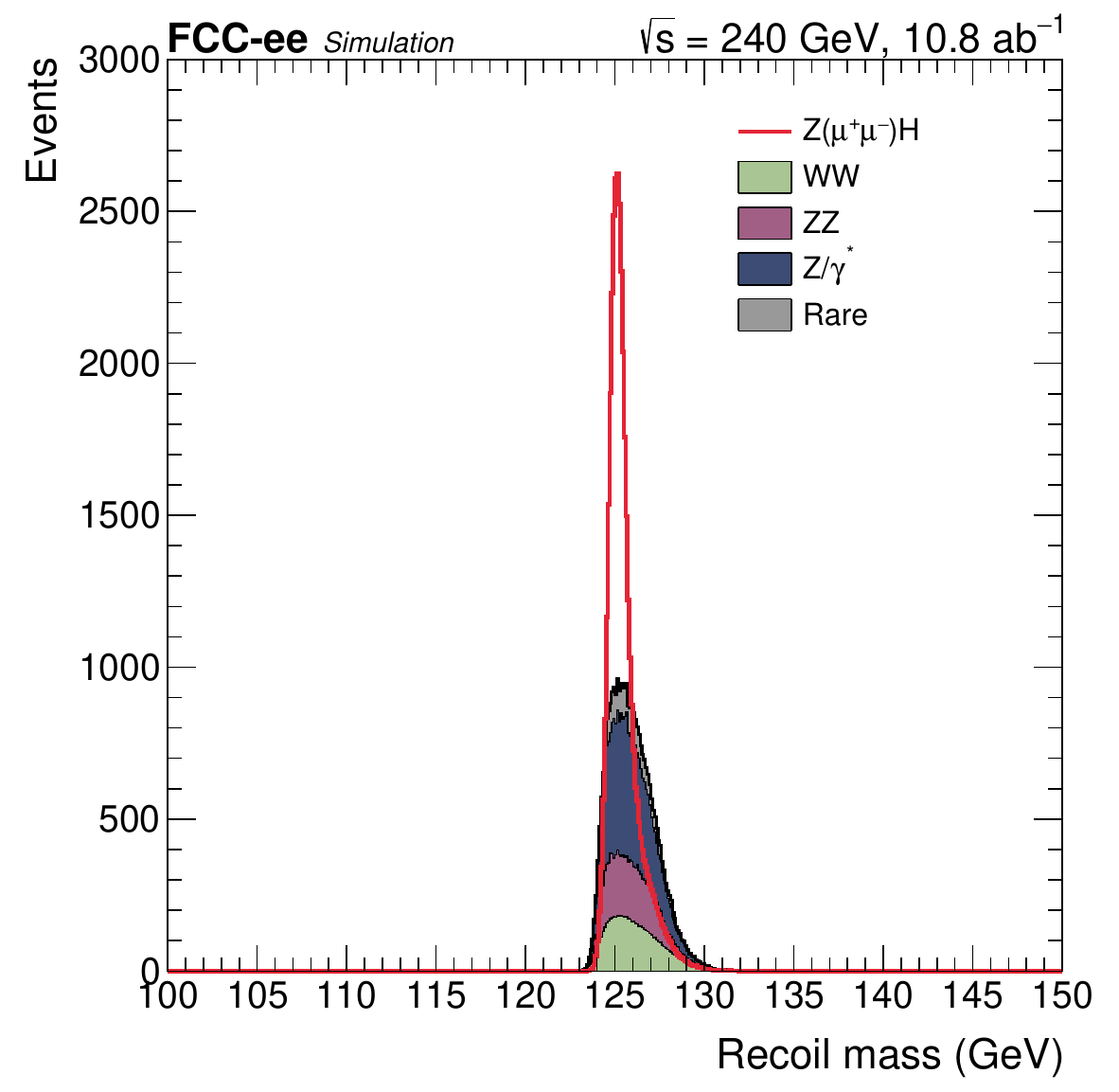}
  \caption{Final fit distributions for the muon final state at \sqrtsZH. The recoil mass is shown for the low (left) and high (right) BDT output regions.}
  \label{fig:sigextr:leptonic_final}
\end{figure}

For the leptonic final states, the fit is performed on the recoil mass distribution in the range $100$--$150$\,GeV, divided into two regions of the BDT output discriminator. This choice keeps the signal extraction tied to the recoil-mass method, while the BDT output is used only to define regions with different signal purity. The lower region, dominated by background, constrains the background normalizations, while the higher region provides the signal sensitivity. The boundary is optimized to maximize the signal significance ($\mathrm{S}/\sqrt{\mathrm{S}+\mathrm{B}}$), yielding thresholds of 0.83 (0.66) for $\mu^+\mu^-$ and 0.88 (0.76) for $\rm e^+e^-$ at $\sqrt{s}=240$ (365)\,GeV. The final templates are shown in Fig.~\ref{fig:sigextr:leptonic_final}.

%Similarly, the hadronic analysis is divided into two regions of the BDT output, with a boundary at 0.75 (0.95) at $\sqrt{s}=240$ (365)\,GeV, chosen to maximize the significance. The higher branching ratio and looser event selection leaves higher event yields available for a two-dimensional template over the jet-pair mass (\mjj) and the recoil mass (\mrec). Including \mjj in the fit allows to constrain the individual background processes and therefore further enhances the sensitivity. 

%Similarly, the hadronic analysis is divided into two regions of the BDT output, with a boundary at 0.75 (0.95) at $\sqrt{s}=240$ (365)\,GeV, chosen to maximize the significance. As in the leptonic channels, the BDT output is used only to define regions with different signal purity. Thanks to the higher branching ratio, looser event selection, and large FCC-ee datasets, sufficient event yields are available to make a two-dimensional template fit in the jet-pair mass (\mjj) and recoil mass (\mrec) possible, thereby providing additional signal sensitivity. The use of \mjj helps constrain the reconstructed $Z\to q\bar q$ candidate and the background composition, while \mrec keeps the signal extraction connected to the recoil-mass method. The projections of the two-dimensional fit distributions onto the jet-pair mass and recoil mass axes, for both the low and high BDT output regions, are shown in Fig.~\ref{fig:sigextr:hadronic_final}.

\begin{figure}[!ht]
  \centering
  \hspace*{-7pt} % align both row of images vertically
  \includegraphics[width=0.49\linewidth]{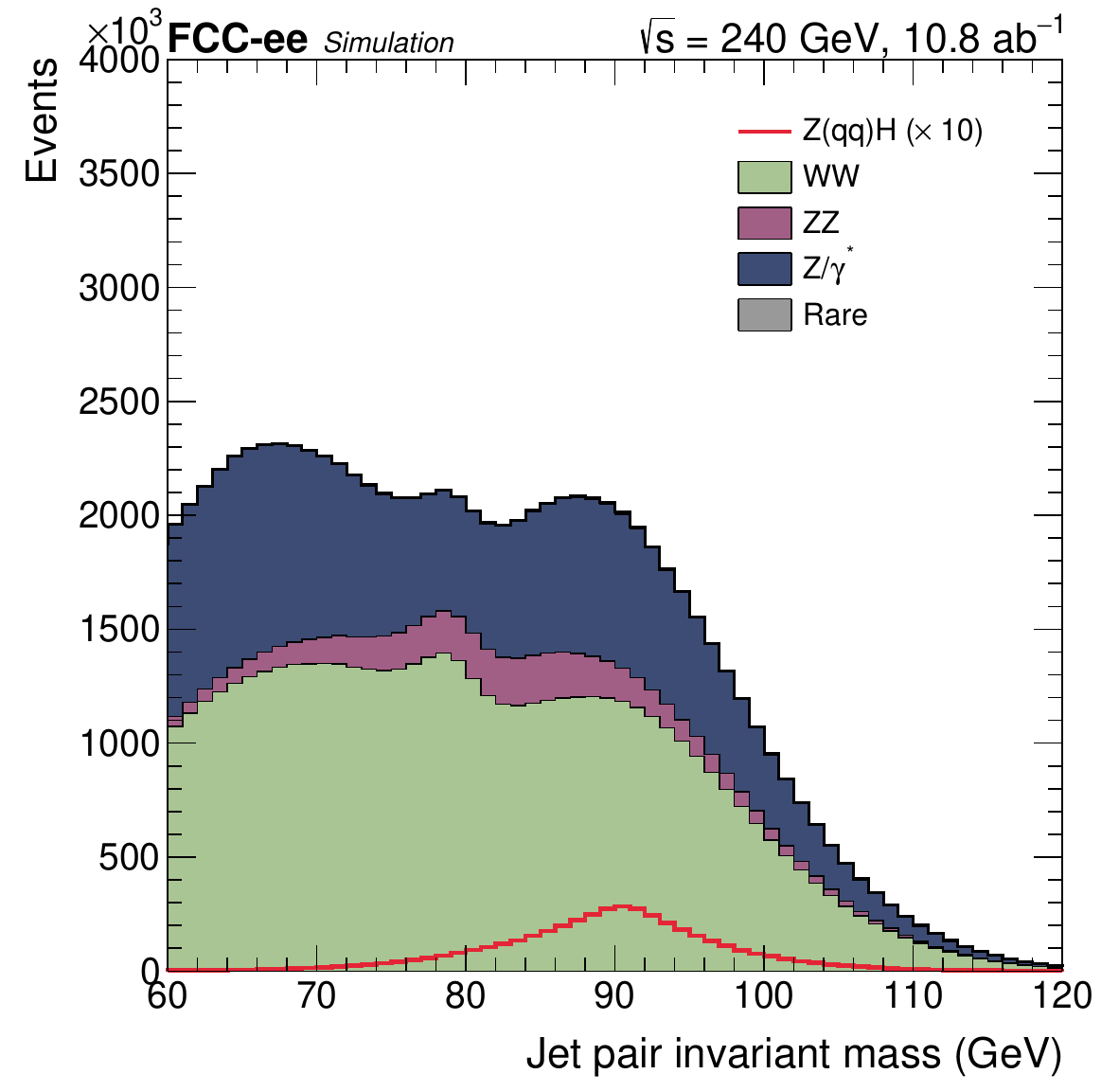}
  \includegraphics[width=0.49\linewidth]{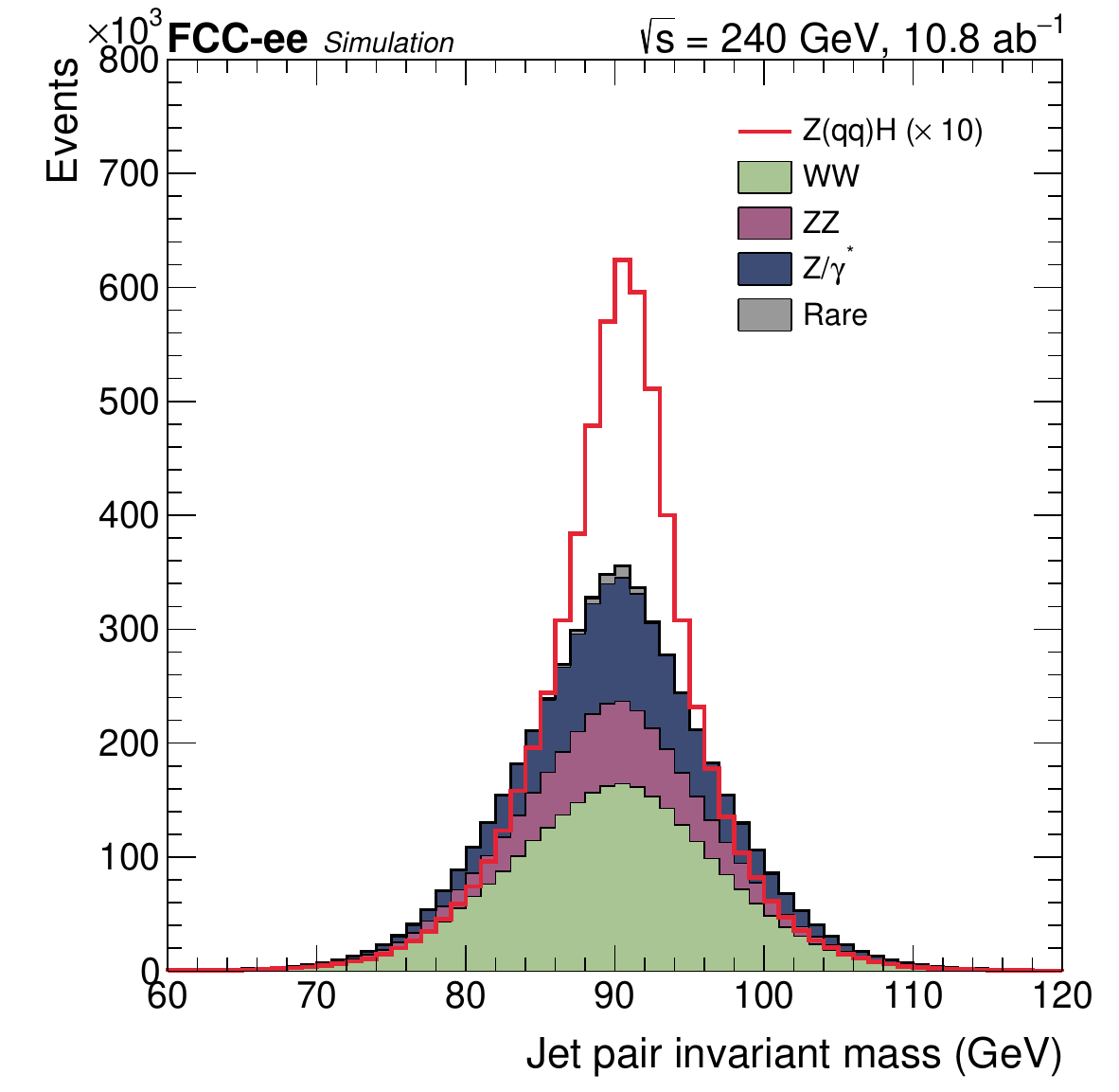}
  \includegraphics[width=0.49\linewidth]{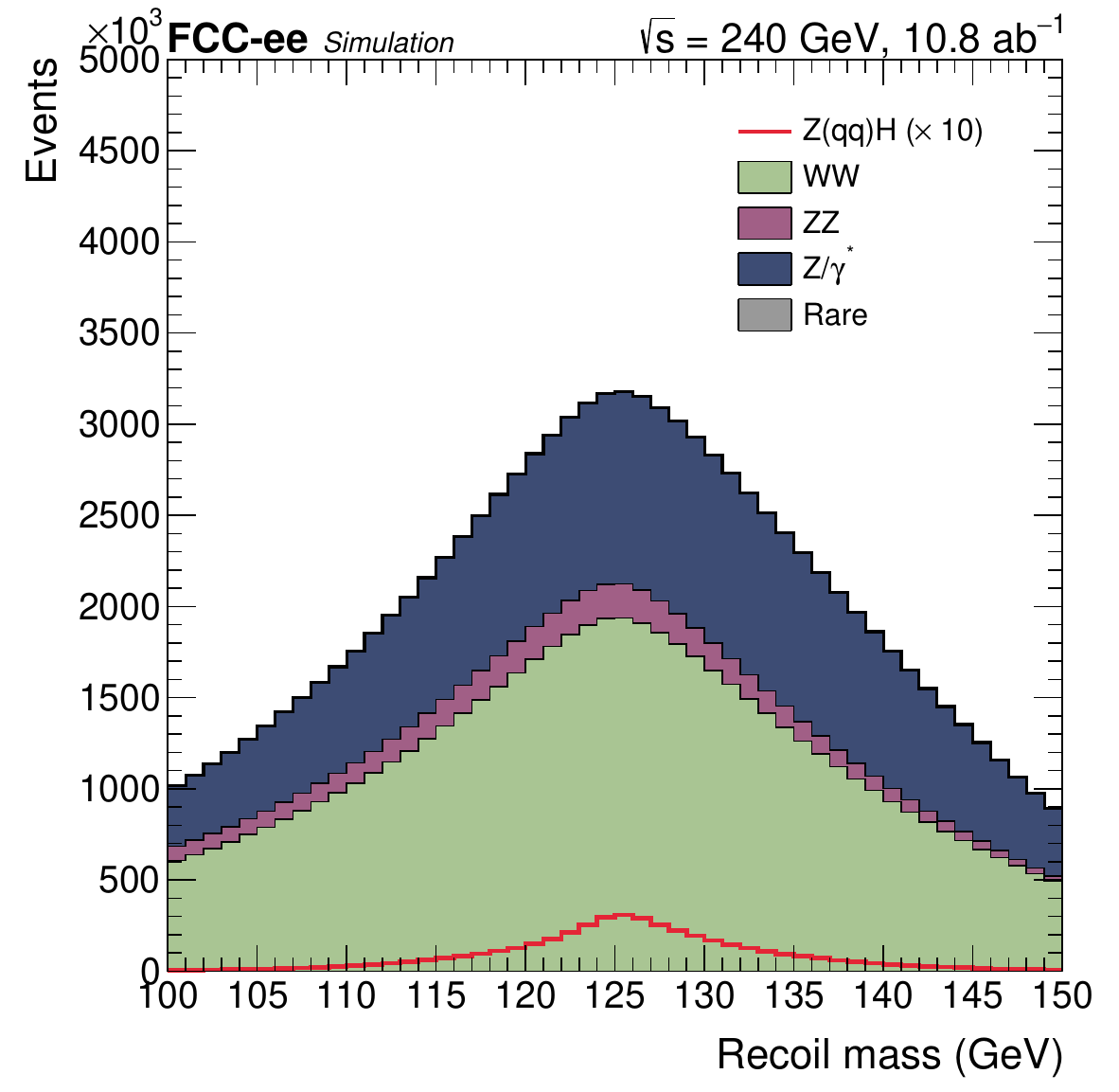}
  \includegraphics[width=0.49\linewidth]{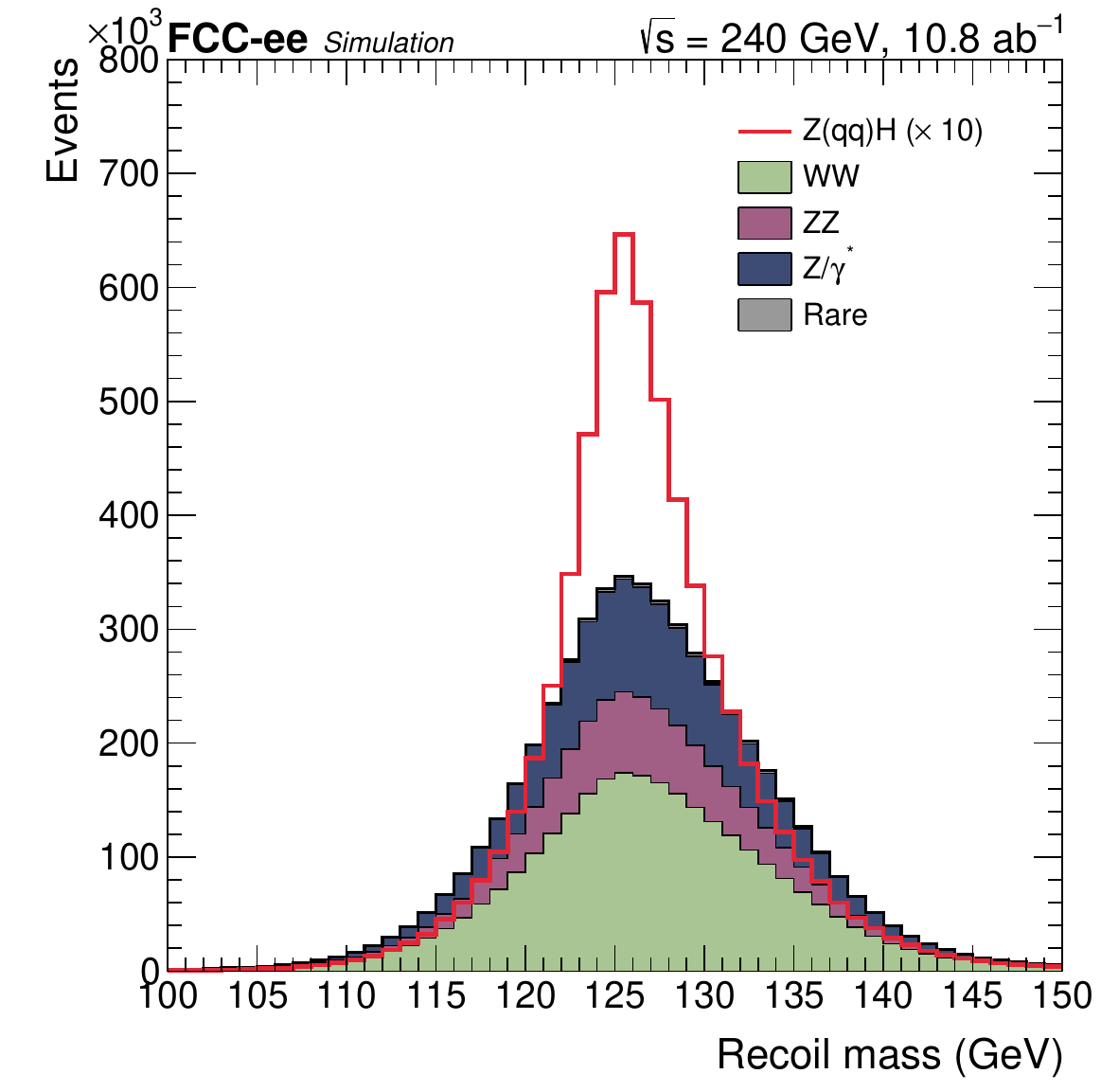}
  \caption{Projections of the two-dimensional final fit distributions for the hadronic final state at \sqrtsZH, onto the jet-pair mass (top) and recoil mass (bottom), shown separately for the low (left) and high (right) BDT output regions.}
  \label{fig:sigextr:hadronic_final}
\end{figure}

Similarly, the hadronic analysis is divided into two regions of the BDT output, with a boundary at 0.75 (0.95) at $\sqrt{s}=240$ (365)\,GeV, chosen to maximize the significance. As in the leptonic channels, the BDT output is used to define regions with different signal purity. The relatively loose pre-selection is designed to retain sizable signal and background yields. Together with the large FCC-ee dataset and the higher branching ratio of $\rm Z\to q\bar q$, this makes a two-dimensional template fit in the jet-pair mass (\mjj) and recoil mass (\mrec) statistically meaningful. The use of \mjj helps constrain the reconstructed $\rm Z\to q\bar q$ candidate and the background composition, while \mrec keeps the signal extraction connected to the recoil-mass method. This strategy improves the background constraints and provides additional signal sensitivity. The projections of the two-dimensional fit distributions onto the jet-pair mass and recoil mass axes, for both the low and high BDT output regions, are shown in Fig.~\ref{fig:sigextr:hadronic_final}.

\subsection{Systematic uncertainties}
\label{subsec:systematics}

The impact of selected experimental systematic effects was evaluated by varying the corresponding detector- or beam-related parameters and repeating the cross-section fit. Among the effects considered, the dominant contribution arises from the beam energy spread (BES). Using abundant $\rm e^+e^- \rightarrow \mu^+\mu^-(\gamma)$ and $\rm e^+e^- \rightarrow e^+e^-(\gamma)$ radiative-return events in the same dataset, the BES can be measured with a precision better than 1\%. Assuming a conservative 1\% uncertainty on the BES, the resulting relative uncertainty on the fitted $\mathrm{ZH}$ cross section is about $0.095\%$ at $\sqrt{s}=\SI{240}{\GeV}$. This effect is mainly due to the impact of the BES on the shape of the recoil mass distribution in the leptonic channels. A similar behavior is observed at $\sqrt{s} = 365$~GeV.

The same radiative-return events are used to constrain the center-of-mass energy and the lepton momentum scale. This sample provides the center-of-mass energy at $240$ and $\SI{365}{\GeV}$ with a precision of a few\MeV, and constrains the momentum scale at the $10^{-5}$ level. Varying these inputs within their assumed uncertainties has a negligible impact on the cross section, at a level below $0.001\%$.
%of $0.001\%$ and below $0.0001\%$, respectively. 
The combined uncertainty from these effects is approximately $0.095\%$, dominated by the BES contribution. 

%The impact of the jet energy scale (JES) and jet energy resolution (JER) has also been evaluated by applying a 1\% shift to the reconstructed jet energies and an additional 1\% Gaussian smearing of the jet energies, added in quadrature to the nominal jet energy resolution of 4\%, respectively. 
The impact of the jet energy scale (JES) and jet energy resolution (JER) has also been evaluated. For the JES, the reconstructed jet energies are shifted by 1\%. For the JER, an additional Gaussian smearing is added in quadrature to the nominal resolution of 4\%, giving $\sqrt{4^2 + 1^2}\,\% \simeq 4.1\%$.
At \sqrtsZH, these variations lead to a relative degradation of the expected precision of the hadronic cross-section measurement by $3.9\%$ and $3.4\%$, respectively. These variations are considered conservative given the large radiative-return and hadronic \WW control samples available for data-driven jet calibration. Furthermore, the quoted results do not include the potential constraints from profiling the JES and JER nuisance parameters in the likelihood fit, which could further reduce their impact. Given their limited impact under these conservative assumptions, JES and JER uncertainties are not included in the final quoted precision. This limited impact contrasts with the leptonic final states, where the signal is concentrated in a recoil-mass peak only a few hundred\MeV wide and a difference of a few $10^{-4}$ in the lepton momentum resolution directly affects the precision. The $\mjj$ and $\mrec$ distributions of the hadronic final state are several\GeV wide and change little under a small variation of the JES or JER. Other detector-related effects, such as selection efficiencies, have not been varied explicitly and would require a dedicated calibration study.

Background processes (\WW, \ZZ, $\mathrm{Z}/\gamma^*$ and rare processes) are modeled as separate components, each with a normalization uncertainty of 1\% included as a nuisance parameter in the fit. This value is an assumption of the present study rather than a measured or calculated uncertainty. It is chosen to be conservative, since these processes will be measured in situ in the same dataset with event yields exceeding the \ZH sample by one to two orders of magnitude. The low-BDT output region and the wide recoil mass range act as control regions that constrain these normalizations directly in the fit, so that their post-fit uncertainties are well below the assumed prior. Because of the event selection and the MVA-based discrimination, the background shapes in the signal region are similar to those of the signal. Shape uncertainties are not included as nuisance parameters in this study, under the assumption that the background shapes will be known at the sub-percent level from theoretical calculations and from dedicated measurements in control regions.

Under the systematic assumptions considered, the expected precision of the \ZH cross-section measurement remains statistically dominated (see Table~\ref{tab:results_xsec}).
%\begin{figure}[h]
%\centering
%\includegraphics[width=0.49\linewidth]{Figures/CrossSection/Fit/xsec_breakDown_combine.pdf}
%\includegraphics[width=0.49\linewidth]{Figures/CrossSection/S365/Fit/xsec_breakDown.pdf}
%\caption{Breakdown of the relative uncertainty on the measured $\mathrm{e^+e^- \to ZH}$ cross section at $\sqrt{s}=240~\mathrm{GeV}$ (left) and $\sqrt{s}=365~\mathrm{GeV}$ (right), showing the considered systematic contributions together with the statistical uncertainty.}
%\label{fig:xsec_Fit_breakdown}
%\end{figure}

\section{Results}\label{sec:results}
The relative uncertainties on the \ZH cross section for the electron, muon, and hadronic final states, and for their combination, are summarized in Table~\ref{tab:results_xsec} for both center-of-mass energies, at the 68\% confidence level. The main results are obtained with the MVA-based configuration, which includes the BDT categorization and, in the hadronic final state, the two-dimensional fit. The combined uncertainty is 0.31\% at \sqrtsZH and 0.52\% at \sqrtsTop.

The hadronic final state provides the highest precision at both energies, owing to its large event yield. Adding the leptonic final states reduces the combined uncertainty by 18\% at \sqrtsZH and by 7\% at \sqrtsTop compared with the hadronic final state alone. The smaller gain at \sqrtsTop reflects the reduced sensitivity of the leptonic final states, caused by the broader recoil-mass distribution shown in Fig.~\ref{fig:xsection:mrecoil}. Among the leptonic final states, \ZmumuH performs best, benefiting from its better resolution and lower background levels. The \ZeeH uncertainty is about 19\% larger at \sqrtsZH, a difference that decreases to about 7\% at \sqrtsTop, where the broader recoil mass distribution reduces the impact of the lepton resolution.

For reference, the results of the cut-based analysis without MVA-based categorization are also reported. In this case, each final state is fitted in a single region, and the hadronic final state is fitted using only the recoil mass distribution. It yields a combined uncertainty of 0.56\% at \sqrtsZH and 1.06\% at \sqrtsTop. The MVA-based analysis therefore improves the combined precision by 45\% at \sqrtsZH and 51\% at \sqrtsTop. The two ingredients contribute unequally. In the leptonic final states, where the fit remains one-dimensional, the categorization alone reduces the uncertainty by 19\% (15\%) at \sqrtsZH (\sqrtsTop). In the hadronic final state, the categorization and the second fit dimension act together and reduce the uncertainty by 68\% (61\%) at \sqrtsZH (\sqrtsTop). This drives the improvement in the combination and reverses the hierarchy at \sqrtsZH, where the hadronic final state is the least precise one in the cut-based configuration.

The statistical-only uncertainties, reported in parentheses in Table~\ref{tab:results_xsec}, are obtained by fixing the background normalization parameters in the likelihood fit. The impact of the background normalization uncertainties is largest in the hadronic final state, as expected from its relatively loose event selection and the substantial remaining background contribution.
For the MVA-based configuration at $\sqrt{s} = \SI{240}{\GeV}$, fixing the background normalizations reduces the uncertainty from 0.52\% to 0.50\% in the leptonic final states, from 0.38\% to 0.30\% in the hadronic final state, and from 0.31\% to 0.26\% in their combination.
This impact is reduced in the MVA-based configuration compared with the cut-based fit, as the low-BDT-score regions provide a background-enriched control region that directly constrains the background normalization parameters. Such constraints are less effective in the cut-based fit, where the signal and background contributions are determined from a single distribution.

%Experimental systematic effects have been evaluated and found to be subdominant with respect to the statistical precision (see Section~\ref{subsec:systematics}), so the measurement is expected to be statistically limited.
The beam- and detector-related effects of Section~\ref{subsec:systematics} (beam energy spread, center-of-mass energy, lepton momentum scale, JES, and JER) are evaluated separately and are not included in the uncertainties of Table~\ref{tab:results_xsec}. They are subdominant with respect to the statistical precision, so the measurement is expected to be statistically limited.

\begin{table}[ht!]
\renewcommand{\arraystretch}{1.1}
\centering
\caption{Relative uncertainties at the 68\% confidence level on the \ZH cross section from the likelihood fit, for the individual final states and their combination. The cut-based configuration fits the recoil mass in a single region per final state. The MVA-based configuration adds the BDT categorization and, for \ZqqH, a second fit dimension. The values in parentheses correspond to the statistical-only uncertainties, obtained with the background normalization parameters fixed.}
\vspace{0.5em}
\label{tab:results_xsec}
{\small
\begin{tabular}{lcccc}
\toprule
 & \multicolumn{4}{c}{$\delta\sigma_{\ZH}/\sigma_{\ZH}$ (\%)} \\
\cmidrule(lr){2-5}
 & \multicolumn{2}{c}{\textbf{\sqrtsZH}} & \multicolumn{2}{c}{\textbf{\sqrtsTop}} \\
\cmidrule(lr){2-3}\cmidrule(lr){4-5}
\textbf{Final state} & \textbf{Cut-based} & \textbf{MVA-based} & \textbf{Cut-based} & \textbf{MVA-based} \\
\midrule
\multicolumn{5}{l}{\textbf{Leptonic final states}} \\
\ZeeH   & 1.01 (0.97) & 0.81 (0.78) & 2.27 (2.16) & 1.91 (1.90) \\
\ZmumuH & 0.84 (0.80) & 0.68 (0.66) & 2.11 (2.03) & 1.79 (1.79) \\
\midrule
\ZllH   & 0.64 (0.62) & 0.52 (0.50) & 1.56 (1.48) & 1.32 (1.30) \\
\midrule
\multicolumn{5}{l}{\textbf{Hadronic final state}} \\
\ZqqH   & 1.20 (0.63) & 0.38 (0.30) & 1.42 (0.63) & 0.56 (0.47) \\
\midrule
\multicolumn{5}{l}{\textbf{Combination}} \\
\ZH     & 0.56 (0.44) & 0.31 (0.26) & 1.06 (0.58) & 0.52 (0.44) \\
\bottomrule
\end{tabular}
}
\end{table}

\section{Assessment of Model Independence}\label{sec:Model-independent}

To assess the degree of model independence of the result, a set of dedicated bias tests is performed. In this context, model independence is defined as the insensitivity of the extracted \ZH cross section to the Standard Model Higgs boson decay modes, such that variations in the relative contributions of individual Higgs boson decay channels do not bias the measurement within the quoted precision. Given that strict model independence cannot be achieved in an absolute sense, this definition provides a practical and testable criterion.

\subsection{Higgs decay mode selection efficiencies}

\begin{figure}[!t]
  \centering
  \includegraphics[width=0.32\linewidth]{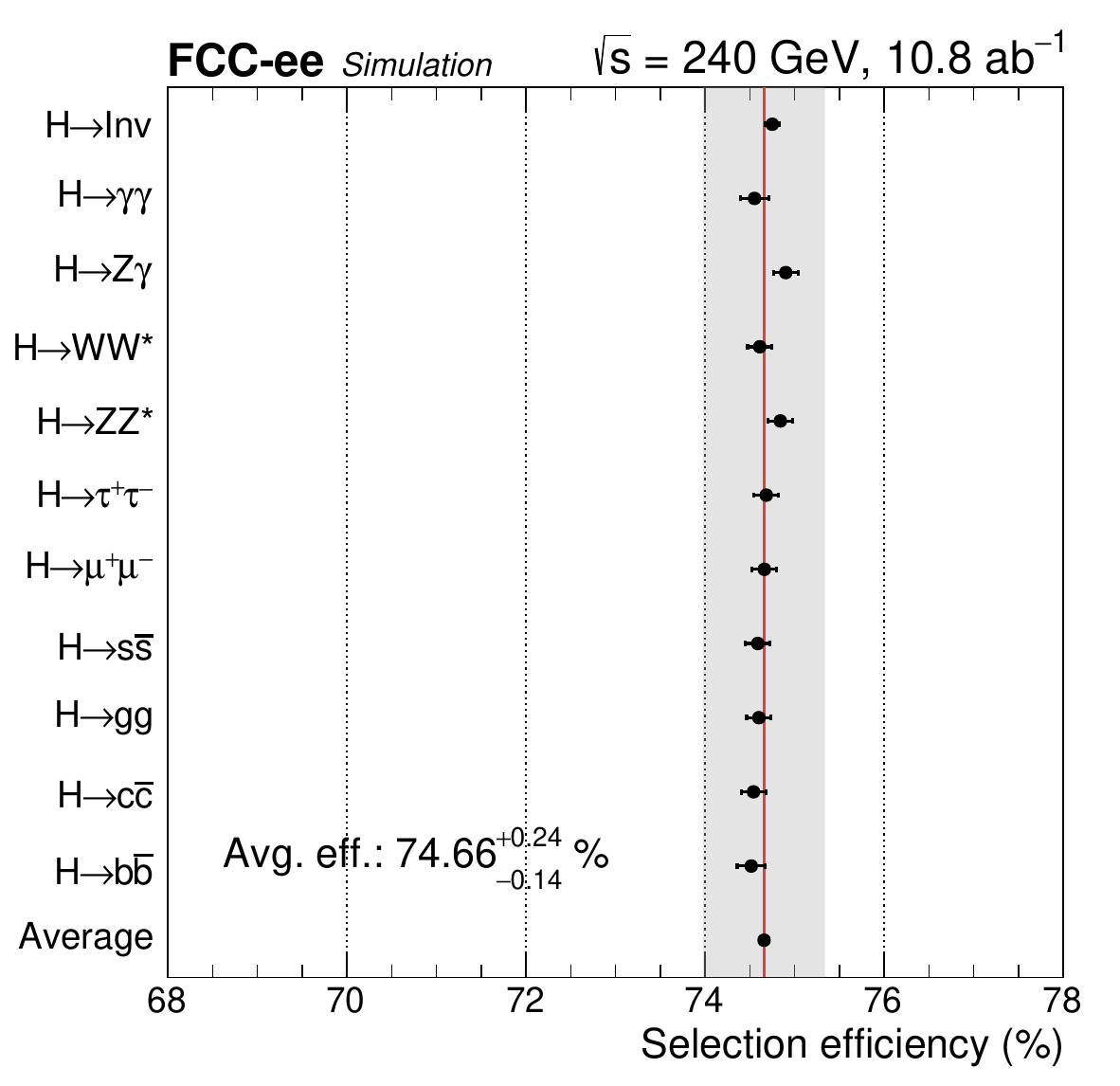}
  \includegraphics[width=0.32\linewidth]{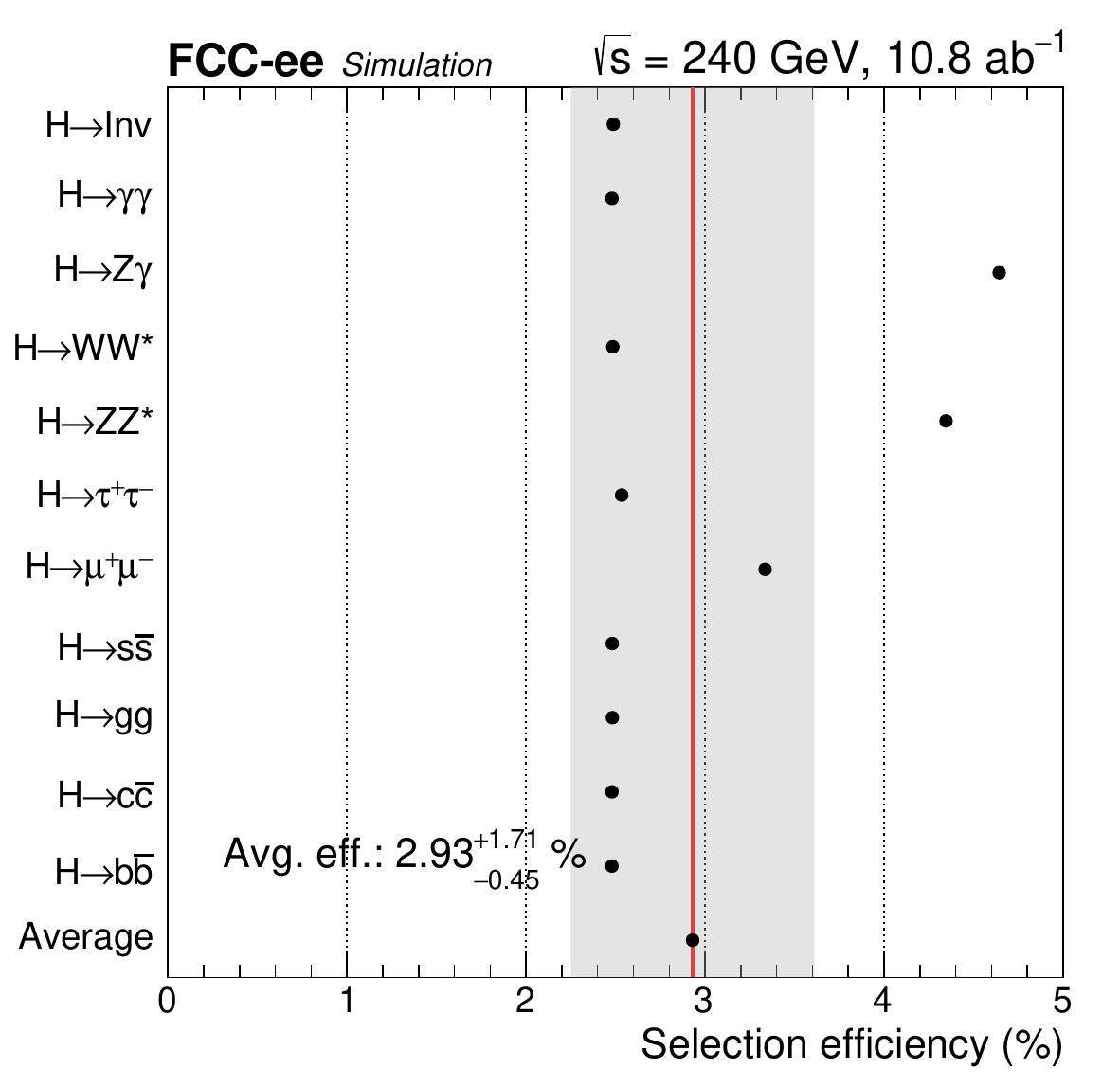}
  \includegraphics[width=0.32\linewidth]{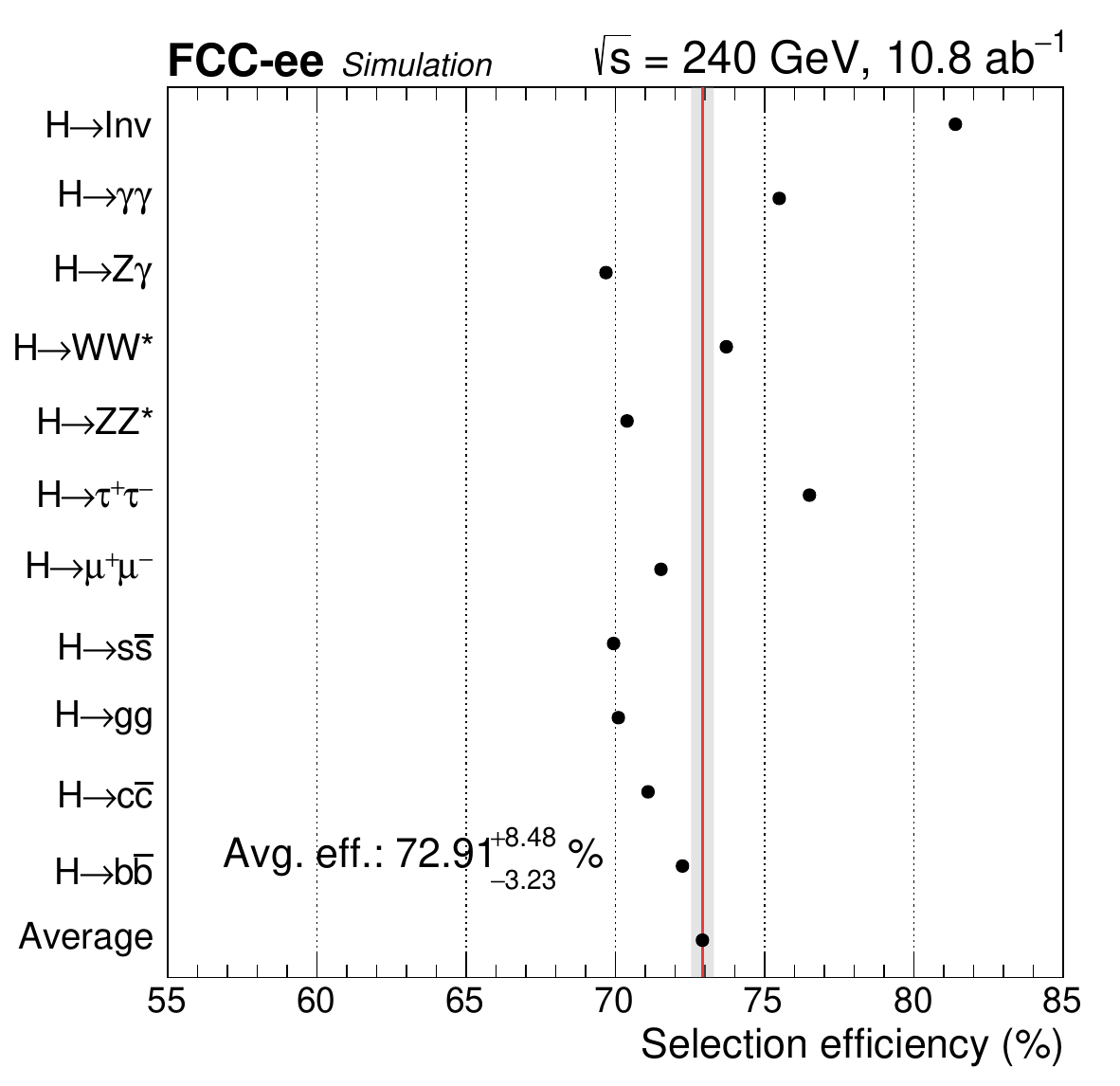}
  \caption{Signal efficiencies after all event selections for the \ZmumuH processes in the muon final state (left), the total \ZH process in the muon final state (middle), and the \ZqqH process in the hadronic final state (right). The gray bands represent the expected uncertainties of the respective final states from Table~\ref{tab:results_xsec}, which include the background normalization uncertainties.}
  \label{fig:seleff}
\end{figure}

By construction, the analysis selection is primarily based on objects that correspond to the decay products of the associated \Z boson. This design choice is intended to minimize sensitivity to the Higgs decay topology. For the leptonic final states, this behavior is validated by examining both the selection efficiencies and the distributions of the BDT output discriminators across different Higgs decay modes. 
The expected uncertainties on the cross section from the fit are $\pm 0.81\%$ for the electron channel and $\pm 0.68\%$ for the muon channel (Table~\ref{tab:results_xsec}), both of which are larger than the observed variations across Higgs decay modes, amounting to ${}^{+0.26}_{-0.29}\%$ and ${}^{+0.24}_{-0.14}\%$, respectively (see Fig.~\ref{fig:seleff}, left). These results indicate that the leptonic selections are insensitive to the Higgs decay mode within the expected uncertainties. In contrast, the hadronic final state does not satisfy this criterion, as the observed variations in selection efficiency across Higgs decay modes significantly exceed the corresponding expected uncertainty (see Fig.~\ref{fig:seleff}, right), indicating an enhanced sensitivity to the Higgs decay products.

However, the requirement that the selection efficiency be independent of the Higgs decay mode alone is not sufficient to guarantee model independence at the level of the final measurement. In particular, \ZH production modes that are not explicitly targeted as reconstructed recoil channels (\ZtautauH and \ZnunuH) can enter the selected phase space through visible decay products of the Higgs boson. As a result, the effective selection efficiency for the inclusive \ZH process can differ from that obtained for the targeted final states. This effect is illustrated in Fig.~\ref{fig:seleff} (middle), which shows the selection efficiency for the inclusive \ZH process in the muon channel. While the average efficiency of $2.93\%$ is dominated by the product of the $\rm Z \to \mu^{+}\mu^{-}$ branching fraction and the $\rm Z(\mu^{+}\mu^{-})H$ selection efficiency, approximately $2.5\%$, matching the lower edge of the observed spread, additional contributions arise from Higgs decay modes such as \Hmumu, $\rm ZZ^{*}$, and $\rm Z\gamma$, which contain genuine muons in the final state. These contributions increase the overall selection efficiency and demonstrate the need for a more comprehensive test of model independence beyond channel-by-channel efficiency comparisons.

\subsection{Bias tests}

As described in Refs.~\cite{Thomson:2015jda,eysermans_2025_jfb44-s0d81}, the test involves perturbing the branching ratio of each Higgs decay mode individually by a fixed amount, such that the resulting variation alters the total \ZH cross section by a relative amount $\rm X$. 
%Due to the relative branching ratios, this perturbation leads to much larger relative changes in certain decay modes (e.g., for $\rm X = 5\%$, the corresponding shift in $\mathcal{B}(\Hgg)$ is approximately 61\%), while $\rm \delta\mathcal{B} \lesssim X$ for dominant modes like \Hbb. 
Since the relative shift in each branching ratio is $X/\mathcal{B}$, the perturbation is amplified for rare decay modes: for $X = 5\%$, the corresponding shift in $\mathcal{B}(\Hgg)$ is approximately 61\%, while the smallest shift, about $8.6\%$, occurs for the dominant \Hbb mode.
The resulting bias in the extracted cross section is then compared to the quoted uncertainty, and the test is considered successful if the bias remains within the quoted uncertainty.

As a first step, bias tests are performed for the muon and electron final states. A conservative variation of $\rm X=5\%$ in the total cross section is applied, serving as a robust benchmark since deviations of this size would be observable and interpretable as signs of new physics. Moreover, the total uncertainty on \sigmaZH, obtained by combining all exclusive decay channels at FCC-ee (quoted as $\sigmaZH \times \mathcal{B}(\Hxx)$), is well below the percent level~\cite{Selvaggi:2025kmd}, further justifying the conservative 5\% variation. The results, summarized in Table~\ref{tab:biastests} at \sqrtsZH, show that all observed biases lie well within the expected uncertainties of the measurement, thereby passing the bias test. Slightly larger biases are observed for \HZa and \HZZ, as these decay modes can also enter the selection through leptonic decays of a \Z boson originating from the Higgs, in the \ZqqH, \ZtautauH, or \ZnunuH final states, which are also treated as part of the signal and thus perturbed in the test. The \Hmumu decay mode shows a smaller bias, appearing only in the muon analysis, as this contribution is largely suppressed by the event selection. These biases are in line with the higher selection efficiencies observed in Fig.~\ref{fig:seleff} (middle).

Bias tests for the hadronic analysis are performed with an injected bias of 1\%, which is justified as the leptonic analyses already measure the total \ZH cross section with a precision well below this level. The results, shown in Table~\ref{tab:biastests} at \sqrtsZH, indicate that all Higgs decay modes pass the test except for \Haa and \Hinv, which exhibit biases slightly larger than $1\sigma$. This arises because these modes exhibit a distinct behavior in the MVA discriminator, as shown in Fig.~\ref{fig:mva:hadronic}. However, given the small branching ratios of these modes, the test is considered conservative and the results remain acceptable.

Bias tests for the full combination at \sqrtsZH, also shown in Table~\ref{tab:biastests}, use the same injected bias of 1\%. Compared to the expected uncertainty of 0.31\%, all Higgs decays pass the test, with only a slight deviation observed for the \Hinv decay mode. Again, this remains acceptable, as the invisible branching ratio is very small, making the test conservative.

To conclude, the observed biases are found to be within the expected uncertainties of the measurement, or to remain at an acceptable level in the conservative tests described above. This confirms the robustness of the total \ZH cross-section determination and demonstrates its model independence within the achieved precision and within the set of Higgs decay modes tested. The corresponding bias tests performed at \sqrtsTop show similar behavior, demonstrating that these conclusions hold at both center-of-mass energies.

%The bias tests presented in this study probe variations among the known Standard Model Higgs decay modes, as well as invisible decays within representative beyond-the-Standard-Model scenarios. By construction, they do not exhaustively cover all possible non-standard Higgs final states, and it remains possible to construct exotic scenarios that populate regions of phase space not explicitly tested. The conclusions of this section therefore apply to the tested Standard Model Higgs decay modes and representative invisible-decay scenarios within the stated level of precision.

The bias tests presented in this study probe variations among the known Standard Model Higgs decay modes, as well as invisible decays within representative beyond-the-Standard-Model scenarios. By construction, they do not exhaustively cover all possible non-standard Higgs final states, and it remains possible to construct exotic scenarios that populate regions of phase space not explicitly tested. Such scenarios would require dedicated studies if they lead to event topologies substantially different from those considered here. The conclusions of this section therefore apply to the tested Standard Model Higgs decay modes and representative invisible-decay scenarios within the stated level of precision.

\begin{table}[!ht]
\renewcommand{\arraystretch}{1.25}
\centering
\caption{Bias tests at \sqrtsZH for the individual final states and their combination. Observed biases are relative and expressed as percentages (\%).}
\vspace{0.6em}
\label{tab:biastests}
{\small
\begin{tabular}{lccccc}
\toprule
\textbf{Decay mode} & \textbf{\ZmumuH} & \textbf{\ZeeH} & \textbf{\ZllH} & \textbf{\ZqqH} & \textbf{Combination} \\
\midrule
\textbf{Injected bias} & 5\% & 5\% & 5\% & 1\% & 1\% \\
\midrule
\Hbb     & $-0.01$ & $+0.00$ & $-0.01$ & $+0.04$ & $+0.02$ \\
\Hcc     & $+0.00$ & $-0.02$ & $-0.01$ & $-0.08$ & $-0.06$ \\
\Hss     & $-0.01$ & $-0.01$ & $-0.01$ & $-0.17$ & $-0.12$ \\
\Hgg     & $+0.00$ & $-0.03$ & $-0.01$ & $-0.04$ & $-0.03$ \\
\Hmumu   & $+0.09$ & $-0.01$ & $+0.04$ & $+0.26$ & $+0.18$ \\
\Htautau & $-0.01$ & $-0.01$ & $-0.01$ & $-0.04$ & $-0.03$ \\
\HZZ     & $+0.34$ & $+0.33$ & $+0.33$ & $+0.02$ & $+0.03$ \\
\HWW     & $-0.01$ & $-0.03$ & $-0.02$ & $-0.07$ & $-0.05$ \\
\HZa     & $+0.36$ & $+0.32$ & $+0.32$ & $-0.17$ & $-0.09$ \\
\Haa     & $+0.00$ & $-0.01$ & $+0.00$ & $+0.43$ & $+0.28$ \\
\Hinv    & $+0.01$ & $+0.01$ & $+0.01$ & $+0.51$ & $+0.34$ \\
\bottomrule
\end{tabular}
}
\end{table}

\section{Conclusion} \label{sec:Conclusion}

In this paper, prospects for model-independent measurements of the \ZH production cross section have been studied at the FCC-ee using the recoil-mass method with simulated data at \sqrtsZHTop and the IDEA detector concept. The analyses include the muon, electron, and hadronic final states of the associated \Z boson, and represent the first consistent treatment of the chosen \Z final states, with a common leptonic selection, an orthogonal hadronic selection, and their robust statistical combination. Event selections are applied to suppress the dominant backgrounds while retaining the signal, and boosted decision trees are employed to further enhance the separation of signal and background in low- and high-score regions, with particular emphasis on minimizing the dependence on the Higgs boson decay mode.

The muon and electron final states are fitted simultaneously using the recoil mass distributions in both regions of the MVA discriminator, achieving a relative precision of 0.52\% (1.32\%) on the \ZH cross section at \sqrtsZH (\SI{365}{\GeV}). The hadronic final state is based on a two-dimensional fit in the \mrec--\mjj plane in two regions of the MVA discriminator, yielding relative precisions of 0.38\% (0.56\%). Combining all three final states results in overall uncertainties of 0.31\% (0.52\%) on the total \ZH production cross section at \sqrtsZH (\SI{365}{\GeV}). These uncertainties are statistically dominated, with the systematic uncertainties contributing below the statistical precision, so that the results scale with the integrated luminosity of the respective run. The model independence of these results has been validated through conservative bias tests, applied consistently across all final states, confirming robustness at the level of the obtained precision.
%\input{Sections/TODO}
%\appendix
%\input{Appendices/app_bdt_perfomance}

\acknowledgments

The work of A. Li is supported by the U.S. Department of Energy, Office of Science, Office of High Energy Physics under contract no. DE-SC0012704.

The work of C. Paus and J. Eysermans is supported by the U.S. Department of Energy, Office of Science, Office of High Energy Physics under contract no. DE‐SC0011939.

We thank our colleague Louis Portales for generating several of the Monte Carlo samples used in these analyses.

% Bibliography

%% [A] Recommended: using JHEP.bst file
\bibliographystyle{JHEP}
\bibliography{FCC}

%% or
%% [B] Manual formatting (see below)
%% (i) We suggest to always provide author, title and journal data or doi:
%% in short all the informations that clearly identify a document.
%% (ii) please avoid comments such as "For a review'', "For some examples",
%% "and references therein" or move them in the text. In general, please leave only references in the bibliography and move all
%% accessory text in footnotes.
%% (iii) Also, please have only one work for each \bibitem.

\end{document}